\documentclass[%
superscriptaddress,
showkeys,
preprint,
 amsmath,amssymb,
 aps,
prc
]{revtex4-1}

\usepackage{graphicx}
\usepackage{dcolumn}
\usepackage{bm}
\usepackage{booktabs}

\usepackage{CJKutf8}
\begin{document}
\begin{CJK*}{UTF8}{gbsn} 

\title{Multimodality of $^{187}$Ir fission studied by Langevin approach}

\author{Y.G. Huang (黄英格)} 
\affiliation{Sino-French Institute of Nuclear Engineering and Technology, Sun Yat-sen University, Zhuhai 519082, China}
\author{F.C. Gu (顾富昌)} 
\affiliation{Sino-French Institute of Nuclear Engineering and Technology, Sun Yat-sen University, Zhuhai 519082, China}
\author{Y.J. Feng (冯玉洁)} 
\affiliation{Sino-French Institute of Nuclear Engineering and Technology, Sun Yat-sen University, Zhuhai 519082, China}
\author{H. Wang (王慧)} 
\affiliation{Sino-French Institute of Nuclear Engineering and Technology, Sun Yat-sen University, Zhuhai 519082, China}
\author{E.X. Xiao (肖尔熙)} 
\affiliation{Sino-French Institute of Nuclear Engineering and Technology, Sun Yat-sen University, Zhuhai 519082, China}
\author{X. Lei (雷昕)} 
\affiliation{Sino-French Institute of Nuclear Engineering and Technology, Sun Yat-sen University, Zhuhai 519082, China}
\author{L. Zhu (祝龙)}
\affiliation{Sino-French Institute of Nuclear Engineering and Technology, Sun Yat-sen University, Zhuhai 519082, China}
\author{J. Su (苏军)}\email{sujun3@mail.sysu.edu.cn} 
\affiliation{Sino-French Institute of Nuclear Engineering and Technology, Sun Yat-sen University, Zhuhai 519082, China}
\affiliation{%
Key Laboratory of Nuclear Data, China Institute of Atomic Energy, Beijing 102413,China
}%

\date{\today}

\begin{abstract}
\begin{description}
\item[Background]
The fission mechanism of sub-lead nuclides remains unclear, especially the types of fission modes involved and their corresponding shell effects.
\item[Purpose]
The aim is to identify the different modes in the fission of $^{187}$Ir, and investigate the corresponding mechanism.
\item[Method]
The three-dimensional Langevin approach considering nucleus elongation, deformation, and mass asymmetry is applied to simulate fission dynamics. The macro-microscopic models are used to calculate the transport coefficients.
\item[Results]
The fragment mass, deformation, and total kinetic energy (TKE) of $^{187}$Ir fission in the excitation energies range from 30 to 45 MeV are calculated.
Based on the mass-TKE correlations, four fission modes are identified, namely two asymmetric standard modes, a symmetric super-long mode, and a symmetric liquid-drop mode.
Strong excitation-energy resistance of two asymmetric modes is found.
The mass distributions show the dominance of single-peak shape, which is in good agreement with experimental data.
The fission potential energy surface and the fission dynamics are analyzed to investigate the origins of the modes and the competition between neutron and proton shell effects.
\item[Conclusions]
Multiple fission modes are included in the $^{187}$Ir fission behind the single-peak-like distribution of observables.
The proton and neutron magic numbers with different asymmetry parameter might heighten the sensitivity to the uncertainties of shell corrections.
\end{description}
\end{abstract}

\maketitle
\end{CJK*}
\section{\label{sec:intro}Introduction}

As a complex quantum many-body problem with significant nuclear deformation, the mechanism of nuclear fission remains incompletely understood since its discovery in 1939.
An essential depiction of nuclear fission envisions the nucleus splitting akin to a liquid drop (LD) \cite{Meitner1939239, Bohr1939426}. 
This illustration falls short in explaining the occurrence of low-energy asymmetric fission in actinides unless the influence of nuclear shell structure is considered, such as through the Strutinsky shell correction method \cite{Strutinsky-98,STRUTINSKY19681}.
Three primary fission modes are then posited to comprehend the fission of actinides, grounded in the neutron shell effects within fragments at $N = 82$ (spherical shell), $N = 88$ (deformed shell), and the LD behavior of the nucleus \cite{BROSA1990167}.
However, further studies report that the mechanism of different modes is still not fully determinate.
A systematic experimental study in 2000 underscored that the average charge of heavy fragments consistently appeared at $Z = 54$, indicating a possible predominant role of the proton shell \cite{SCHMIDT2000221}.
A microscopic calculation in 2018 suggested that the proton shell with octupole deformation might be accountable for the two asymmetric fission modes \cite{scamps2018impact}.
Through an analysis of the asymmetry modes in the fission of actinides and preactinides, the proton shell emerged as a notable link in fission properties between these two nuclide regions \cite{MAHATA2022136859}.
The application of Bayesian neural networks in predicting fission charge distribution demonstrated its potential in data evaluation, potentially aiding in estimating inconsistent modes \cite{PhysRevC.103.034621}.
Since the multimodality of the fission process is a crucial theoretical concept for nuclear data \cite{international2008iaea}, definitively determining the fission modes and their corresponding shell effects is imperative for both fission theory and nuclear data evaluation \cite{MAHATA2022136859}.

In 2010, the discovery of asymmetric fission in $^{180}$Hg expanded the realm of asymmetric fission, but challenged the prior understandings of fission dynamics \cite{PhysRevLett.105.252502}.
Various theories have uncovered the intricate potential energy landscape and shell effects of sub-lead nuclides, presenting a heightened challenge for intuitive anticipation compared to actinides \cite{PhysRevC.85.024306, PhysRevC.90.021302, PhysRevC.100.041602, ichikawa2019microscopic, PhysRevC.106.024307, bernard2023hartree}.
To comprehend the fission properties of sub-lead nuclides, a key question revolves around identifying different modes and their corresponding shell effects. 
However, the high excitation energy in experiments typically weakens the shell effects, hindering the direct identification of mode components. 
This necessitates an analysis that connects multiple observables in both theory and experiment.
The correlation between fragment mass, deformation, and total kinetic energy (TKE) is employed to identify fission modes in the actinides and trans-actinide region \cite{usang2018}. 
Recently, a two-dimensional fitting approach, based on fragment mass and TKE, has been developed, enabling a rigorous distinction of three modes in $^{178}$Pt fission \cite{SWINTONBLAND2023137655}.

The fission mechanism of $^{187}$Ir presents a challenge in the study of sub-lead nuclide fission. Several theories predict that asymmetric modes dominate for $^{187}$Ir \cite{PhysRevC.93.034620, PhysRevC.91.044316, SCHMIDT2016107}, contradicting early experimental findings \cite{itkis1991}. Notably, there are significant discrepancies among different theoretical results.
Recent measurements by Dhuri \textit{et al.} on $^{187}$Ir fission via fusion-fission reactions verify the single-peak mass yields of fragments. 
Further discussions are then required to address the discrepancies between theory and experiment \cite{PhysRevC.106.014616}.
As a common theoretical method, the Langevin approach has been widely used in the nuclear fission dynamics studies and the fission observables simulations,
and is recently developed to higher dimension \cite{PhysRevC.96.064616,ivanyuk20235dimensional}.
In this study, the three-dimensional Langevin approach is applied to investigate the fission of $^{187}$Ir. 
Fission modes are identified based on mass-TKE correlations of fragments, and the corresponding mechanisms are discussed.

This paper is organized as follows.
In Sec.~\ref{sec:model}, we present the theoretical framework.
Related details could also be seen in our previous work \cite{PhysRevC.106.054606}.
In Sec.~\ref{sec:results}, the calculated results as well as some discussions are given.
In Sec.~\ref{sec:summary}, a summary of our work are presented.

\section{\label{sec:model}Theoretical framework}
\subsection{\label{shape}Nuclear shape parametrization}
In this work, the $z$-axis-rotational symmetric parametrization of nuclear shape in the deformed two-center shell model (TCSM) \cite{MaruhnGreiner-96} is used.
In the cylindrical coordinate, the nuclear radius $\rho_s$ is expressed as
\begin{equation}
\rho ^2_{\mathrm{s}}=\begin{cases}
	b_{1}^{2}\left( 1-\frac{{z_{1}^{\prime}}^2}{a_{1}^{2}} \right)&		z<z_1,\\
	b_{1}^{2}\left( 1-\frac{{z_{1}^{\prime}}^2}{a_{1}^{2}}\left( 1+c_1z_{1}^{\prime}+d_1{z_{1}^{\prime}}^2 \right) \right) /\left( 1+g_1{z_{1}^{\prime}}^2 \right)&		z_1 \leqslant z < 0,\\
	b_{2}^{2}\left( 1-\frac{{z_{2}^{\prime}}^2}{a_{2}^{2}}\left( 1+c_2z_{2}^{\prime}+d_2{z_{2}^{\prime}}^2 \right) \right) /\left( 1+g_2{z_{2}^{\prime}}^2 \right)&		0 \leqslant z<z_2,\\
	b_{2}^{2}\left( 1-\frac{{z_{2}^{\prime}}^2}{a_{2}^{2}} \right)&		z_2 \leqslant z,\\
\end{cases}
\label{eq:shape}
\end{equation}
where $z_{i}^{\prime}=z-z_i$, $z_i$ is the fragment centers.
$a_i$ and $b_i$ are the short and long axes of fragment, the subscripts $i=1,2$ represent the two fission fragments.
Three nuclear deformation parameters are free.
Namely the dimensionless elongation of nucleus $z_0/R_0$ with $z_0=z_2-z_1$, the mass asymmetry $\eta =\left( A_1-A_2 \right) /\left( A_1+A_2 \right)$,
and the deformation $\delta _i=\left( 3\beta _i-3 \right) /\left( 1+2\beta _i \right)$ with $\beta _i=a_i/b_i$.
$R_0=r_0A_{CN}^{1/3}$ denotes the radius of spherical compound nucleus.
The deformation of two fragments is considered to be equal, i.e., $\delta _1=\delta _2=\delta$.
According to the position of the saddle point shown in Fig.~\ref{fig:initial}(a),
the neck parameter $\epsilon$ is fixed at 0.24,
which is defined as the ratio of the actual potential to the deformed oscillator potential along the symmetry axis at $z=0$.
This three-dimensional parametrization can be generalized to four dimensions by considering $\delta_1$ and $\delta_2$ as two degrees of freedom \cite{PhysRevC.96.064616}, and to five dimensions by unfixing the $\epsilon$ \cite{ivanyuk20235dimensional}.

\subsection{\label{sec:langevin}The Langevin approach}
The multidimensional Langevin equations are written as follow:
\begin{equation}
\frac{dq_i}{dt}=\left( m^{-1} \right) _{ij}p_j,
\label{eq:langevin1}
\end{equation}
\begin{equation}
\frac{dp_i}{dt}=-\frac{\partial V}{\partial q_i}-\frac{1}{2}\frac{\partial \left( m^{-1} \right) _{jk}}{\partial q_i}p_jp_k-\gamma _{ij}\left( m^{-1} \right) _{jk}p_k+g_{ij}\Gamma _j\left( t \right),
\label{langevin2}
\end{equation}
where $\boldsymbol{q}=\left\{ z_0/R_0,\delta ,\eta \right\}$, $\boldsymbol{p}$, $\boldsymbol{m}^{-1}$, $\boldsymbol{\gamma}$ and $\boldsymbol{g}$ are the generalized coordinate, the generalized momentum, the inverse of the inertia tensor, the friction tensor, and the random force strength, respectively.
$\boldsymbol{\Gamma}$ is the normalized Gaussian random force. 
In Eqs.~(\ref{eq:langevin1}) and (\ref{langevin2}) and the following equations, the Einstein summation convention over equal pair indices is used.
The random force strength is related to the friction by the fluctuation-dissipation theorem, i.e.
\begin{equation}
g_{ik}g_{jk}=\gamma _{ij}T,
\end{equation}
where $T$ is the nuclear temperature.
According to the Fermi gas model, the nuclear temperature can be calculated by $E_{\mathrm{int}}=a_nT^2$ with the level density parameter $a_n=A_{\mathrm{CN}}/12$. $E_{\mathrm{int}}$ is the intrinsic excitation energy of compound nucleus related to total excitation energy $E^*$ as
\begin{equation}
E_{\mathrm{int}}\left( \boldsymbol{q},t \right) =E^*-\frac{1}{2}\left( m^{-1} \right) _{ij}p_ip_j-V\left( \boldsymbol{q},T=0 \right),
\label{eq:energy}
\end{equation}
where $V$ is the fission potential energy.
The $T$ will be updated once the $E_{\mathrm{int}}$ is adjusted at each time step.
When the nucleus reaches scission point, the TKE of fragments is calculated as \cite{AritomoChiba-40}
\begin{equation}
\mathrm{TKE}=V_\mathrm{Coul}+E_\mathrm{kin},
\end{equation}
where $V_\mathrm{Coul}=Z_1Z_2e^2/D$ and $E_{\mathrm{kin}}=1/2\left( m^{-1} \right) _{z_0/R_0,z_0/R_0}p_{z_0/R_0}^{2}$ are the Coulomb repulsion energy of fragments, and the kinetic energy of nuclear collective motion in the fission direction at scission point.
$D$ is the fragments charge center distance.

\subsection{\label{sec:pes}Potential energy}
Under the framework of the macro-microscopic model, the fission potential energy is calculated as \cite{ignatyuk1979}
\begin{equation}
V(\boldsymbol{q}, T)=E_{\mathrm{mac}}(\boldsymbol{q})+E_{\mathrm{shell}}(\boldsymbol{q}, T=0)\Phi(T),
\end{equation}

\begin{equation}
\label{eq:phi}
\Phi(T)=\mathrm{exp}(-a_nT^2/E_d),
\end{equation}
where the damping parameter of shell correction $E_d=28$ MeV, chosen by comparing the calculated and experimental fragments mass distributions at low excitation energies.
The shell correction energy $E_{\mathrm{shell}}$ is given by the Strutinsky shell correction method \cite{Strutinsky-98,STRUTINSKY19681}.
The nuclear single-particle levels at each deformation grid point are calculated by the TCSM \cite{MaruhnGreiner-96}.
One TCSM program is available on the NRV web knowledge base \cite{KARPOV2017112}.
For macroscopic energy, only the nuclear surface energy $E_{n}$ and the Coulomb energy $E_{\mathrm{C}}$ change during fission due to the nuclear volume conservation assumption.
By setting the potential of spherical nucleus as the potential origin, the macroscopic part is calculated as
\begin{equation}
E_{\mathrm{mac}}=E_{n}-E_{n0}+E_{\mathrm{C}}-E_{\mathrm{C}0}.
\end{equation}
Based on the finite range liquid drop model (FRLDM),
the nuclear surface and Coulomb energy can be written in triple integral form as \cite{PhysRevC.20.992,DaviesNix-31}
\begin{equation}
\begin{split}
E_{n}&=\frac{a_{\mathrm{S}}\left( 1-k_{\mathrm{S}}I^2 \right)}{4\pi r_{0}^{2}}\iiint{\begin{array}{c}
	\left\{ 2-\left[ \left( \frac{\sigma}{a} \right) ^2+2\frac{\sigma}{a}+2 \right] e^{-\sigma /a} \right\}\\
\end{array}}
\\
&\times \rho _{\mathrm{s}}\left( z \right) \left[ \rho _{\mathrm{s}}\left( z \right) -\rho _{\mathrm{s}}\left( z^{\prime} \right) \cos \phi -\frac{d\rho _{\mathrm{s}}\left( z \right)}{dz}\left( z-z^{\prime} \right) \right]
\\
&\times \rho _{\mathrm{s}}\left( z^{\prime} \right) \left[ \rho _{\mathrm{s}}\left( z^{\prime} \right) -\rho _{\mathrm{s}}\left( z \right) \cos \phi -\frac{d\rho _{\mathrm{s}}\left( z^{\prime} \right)}{dz^{\prime}}\left( z^{\prime}-z \right) \right] \frac{dzdz^{\prime}d\phi}{\sigma ^4},
\end{split}
\end{equation}
\begin{equation}
\begin{split}
E_{\mathrm{C}}&=\pi \rho _{0}^{2}\iiint{}
	\left\{ \frac{\sigma ^3}{6}-a_{\mathrm{d}}^{3}\left[ 2\frac{\sigma}{a_{\mathrm{d}}}-5+\left( 5+3\frac{\sigma}{a_{\mathrm{d}}}+\frac{1}{2}\left( \frac{\sigma}{a_{\mathrm{d}}} \right) ^2 \right) e^{-\sigma /a_{\mathrm{d}}} \right] \right\}
\\
&\times \rho _{\mathrm{s}}\left( z \right) \left[ \rho _{\mathrm{s}}\left( z \right) -\rho _{\mathrm{s}}\left( z^{\prime} \right) \cos \phi -\frac{d\rho _{\mathrm{s}}\left( z \right)}{dz}\left( z-z^{\prime} \right) \right]
\\
&\times \rho _{\mathrm{s}}\left( z^{\prime} \right) \left[ \rho _{\mathrm{s}}\left( z^{\prime} \right) -\rho _{\mathrm{s}}\left( z \right) \cos \phi -\frac{d\rho _{\mathrm{s}}\left( z^{\prime} \right)}{dz^{\prime}}\left( z^{\prime}-z \right) \right] \frac{dzdz^{\prime}d\phi}{\sigma ^4},
\end{split}
\end{equation}
where the integration domain is the full space and the $\sigma$ can be derived by the coordinate system transformation as
\begin{equation}
\begin{split}
\sigma= [ \rho _{\mathrm{s}}^{2}\left( z \right) +\rho _{\mathrm{s}}^{2}\left( z^{\prime} \right) -2\rho _{\mathrm{s}}^{}\left( z \right) \rho _{\mathrm{s}}^{}\left( z^{\prime} \right) \cos \phi +z^2+{z^{\prime}}^2-2zz^{\prime} ] ^{1/2}.
\end{split}
\end{equation}
The potential energy at every general coordinate is calculated to build the fission potential energy surface (PES) and save for solving the Langevin equation.

\begin{figure}
\includegraphics[width=8.6cm]{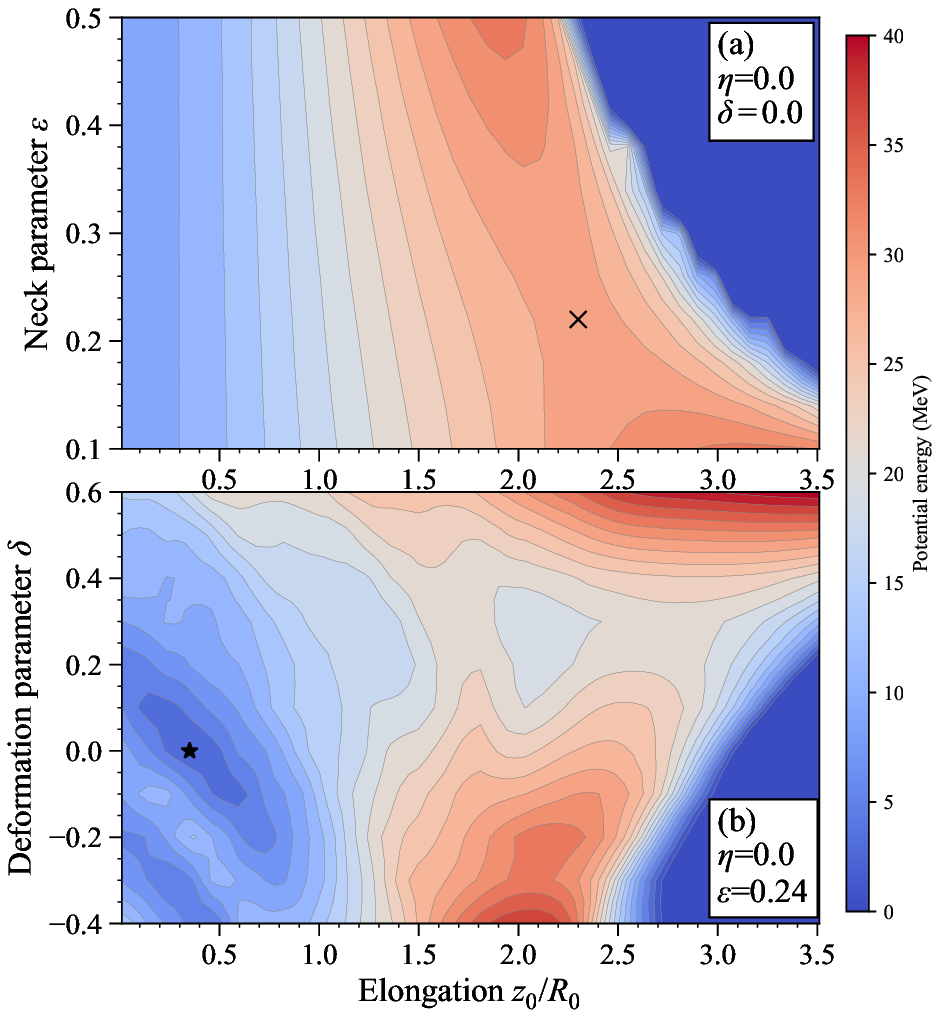}
\caption{\label{fig:initial}(a) The macroscopic potential energy surface at $\eta=0$ and $\delta=0$.
The cross is the saddle point.
(b) The potential energy surface at $\eta=0$ and $\epsilon=0.24$.
The star is the initial position of Langevin trajectories.
}
\end{figure}

To solve the Langevin equation, the initial momentum and coordinate are required for iteration.
The initial momentum is set to zero.
The initial coordinate is set as $\{z_0/R_0,~\delta,~\eta\} = \{0.35,~0,~0\}$, which is the local potential minimum near ground state, as shown in Fig.~\ref{fig:initial}(b).
The limit of the $\{z_0/R_0,~\delta,~\eta\}$ is $\{[0.0,3.6],~[-0.4,0.6],~[-0.4,0.4]\}$.
Once the trajectory reaches the coordinate boundaries, the Langevin calculation will restart from the initial state.
Due to the potential minimum, the trajectory is often trapped and would not cross over the potential barrier.
For reducing the computing resource consumption, some artificial measures could be applied to restrict the trajectory since we assume that the nucleus has a tendency to fission \cite{mavlitov1992combining}.
In the present work, the Gaussian distribution center of the random force at $z_0/R_0$ direction is set at $0.1\sigma_{\mathrm{Gaussian}}$ with $\sigma_{\mathrm{Gaussian}}=\sqrt{2}$.
The centers at $\eta$ and $\delta$ directions keep $0$ for all calculations.
The Langevin trajectory would end once it reaches the scission point,
where the nuclear neck radius is less than 2 fm chosen by comparing the calculated and experimental $\left< \mathrm{TKE} \right>$ at low excitation energies.

\subsection{\label{sec:tensor}Inertia and friction tensor}
Under the Werner-Wheeler approximation, the inertia tensor is calculated by the hydrodynamic model \cite{PhysRevC.13.2385}
\begin{equation}
m_{ij}\left( \boldsymbol{q} \right) =\pi \rho _m\int_{z_{\min}}^{z_{\max}}{\rho _{\mathrm{s}}^{2}\left( z,\boldsymbol{q} \right) \left[ A_iA_j+\frac{1}{8}\rho _{\mathrm{s}}^{2}\left( z,\boldsymbol{q} \right) A_{i}^{'}A_{j}^{'} \right] dz},
\end{equation}
\begin{equation}
\label{eq:Ai}
A_i=\frac{1}{\rho _{\mathrm{s}}^{2}\left( z,\boldsymbol{q} \right)}\frac{\partial}{\partial q_i}\int_z^{z_{\max}}{\rho _{\mathrm{s}}^{2}\left( z^{'},\boldsymbol{q} \right) dz^{'}}-\frac{\partial z_{\mathrm{cm}}}{\partial q_i},
\end{equation}
where the nucleus density $\rho _m=1.668\times 10^{-45}\,\,\mathrm{MeV}\mathrm{s}^2\mathrm{fm}^{-5}$.
$A_{i}^{\prime}$ is the derivative of $A_i$ with respect to $z$.
Since the nuclear center-of-mass coordinate $z_{\mathrm{c.m.}}$ changes during fission process, the derivative term $\partial z_{\mathrm{c.m.}}/\partial q_i$ is subtracted in Eq.~(\ref{eq:Ai}) to exclude the spurious contribution of $z_{\mathrm{c.m.}}$ \cite{Krappe2012TheoryON,ivanyuk20235dimensional}.
The friction tensor is calculated by macroscopic one-body wall-and-window model.
For neckless nucleus, the wall dissipation is given as \cite{BlockiBoneh-22,SierkNix-51}
\begin{equation}
\begin{split}
\gamma _{ij}^{\mathrm{Wall}}\left( \boldsymbol{q} \right) =\frac{1}{2}\pi \rho _m\bar{v}\int_{z_{\min}}^{z_{\max}}{\left( \frac{\partial \rho _{\mathrm{s}}^{2}}{\partial q_i}+\frac{\partial \rho _{\mathrm{s}}^{2}}{\partial z}\frac{\partial z_{\mathrm{c.m.}}}{\partial q_i} \right) \left( \frac{\partial \rho _{\mathrm{s}}^{2}}{\partial q_j}+\frac{\partial \rho _{\mathrm{s}}^{2}}{\partial z}\frac{\partial z_{\mathrm{c.m.}}}{\partial q_j} \right) \left[ \rho _{\mathrm{s}}^{2}+\frac{1}{4}\left( \frac{\partial \rho _{\mathrm{s}}^{2}}{\partial z} \right) ^2 \right] ^{-1/2}dz},
\end{split}
\end{equation}
where $\bar{v}=3v_f/4\approx 6.4\times 10^{22}~\mathrm{fm}/\mathrm{s}$ is the average velocity of nucleons based on Fermi velocity.
The terms proportional to $\partial z_{\mathrm{c.m.}}/\partial q_i$ are added for the same purpose of excluding the spurious contribution of $z_{\mathrm{c.m.}}$.
When the neck of nucleus appeared, the friction tensor is calculated as \cite{AdeevVanin-53}
\begin{equation}
\gamma _{ij}^{\mathrm{W}+\mathrm{W}}\left( \boldsymbol{q} \right) =\gamma _{ij}^{\mathrm{Wall}2}+\gamma _{ij}^{\mathrm{Window}}
\end{equation}
with
\begin{equation}
\gamma _{ij}^{\mathrm{Wall}2}\left( \boldsymbol{q} \right) =\frac{1}{2}\pi \rho _m\bar{v}\left( \int_{z_{min}}^{z_N}{I_L\left( z \right) dz}+\int_{z_N}^{z_{min}}{I_R\left( z \right) dz} \right) ,
\end{equation}
\begin{equation}
\begin{split}
I_v=\left( \frac{\partial \rho _{\mathrm{s}}^{2}}{\partial q_i}+\frac{\partial \rho _{\mathrm{s}}^{2}}{\partial z}\frac{\partial D_v}{\partial q_i} \right) \left( \frac{\partial \rho _{\mathrm{s}}^{2}}{\partial q_j}+\frac{\partial \rho _{\mathrm{s}}^{2}}{\partial z}\frac{\partial D_v}{\partial q_j} \right)
\times \left[ \rho _{\mathrm{s}}^{2}+\frac{1}{4}\left( \frac{\partial \rho _{\mathrm{s}}^{2}}{\partial z} \right) ^2 \right] ^{-1/2},
\end{split}
\end{equation}
where $v=L, R$ represents the prefragments on the left and right sides of the neck.
$z_N$ is the position of the smallest neck radius, which equals 0 in the present work.
$D_v$ is the mass-center position of two prefragments.
The window dissipation is calculated as \cite{Swiatecki-52}
\begin{equation}
\gamma _{ij}^{\mathrm{Window}}\left( \boldsymbol{q} \right) =\frac{1}{2}\rho _m\bar{v}\left( \frac{\partial R_{12}}{\partial q_i}\frac{\partial R_{12}}{\partial q_j}\Delta \sigma +\frac{32}{9\Delta \sigma}\frac{\partial V_{\mathrm{R}}}{\partial q_i}\frac{\partial V_{\mathrm{R}}}{\partial q_j} \right) ,
\end{equation}
where $R_{12}$ denotes the distance between the mass center of prefragments.
$\Delta \sigma$ represents the window area.
$V_R$ is the volume of right prefragments.
Nix and Sierk proposed a phenomenological formula to smoothly transition this two types of friction \cite{RayfordNixSierk-35}
\begin{equation}
\gamma _{ij}=\tau \left( \gamma _{ij}^{\mathrm{W}+\mathrm{W}} \right) +\left( 1-\tau \right) \gamma _{ij}^{\mathrm{Wall}}.
\end{equation}
The choice of $\tau$ is subjective \cite{Bzocki-37,Feldmeier-36}.
The expression in Ref.~\cite{RayfordNixSierk-35} is used in the present work
\begin{equation}
\tau =\cos ^2\left( \frac{\pi}{2}\frac{r_{N}^{2}}{b_{\min}^{2}} \right) ,  b_{\min}=\min \left( b_1, b_2 \right) ,
\end{equation}
where $r_N$ is the neck radius,
$b_1$ and $b_2$ are the long axes of prefragment in Eq.~(\ref{eq:shape}).

\section{\label{sec:results}Results and discussion}


\begin{figure*}
\includegraphics[width=16.5cm]{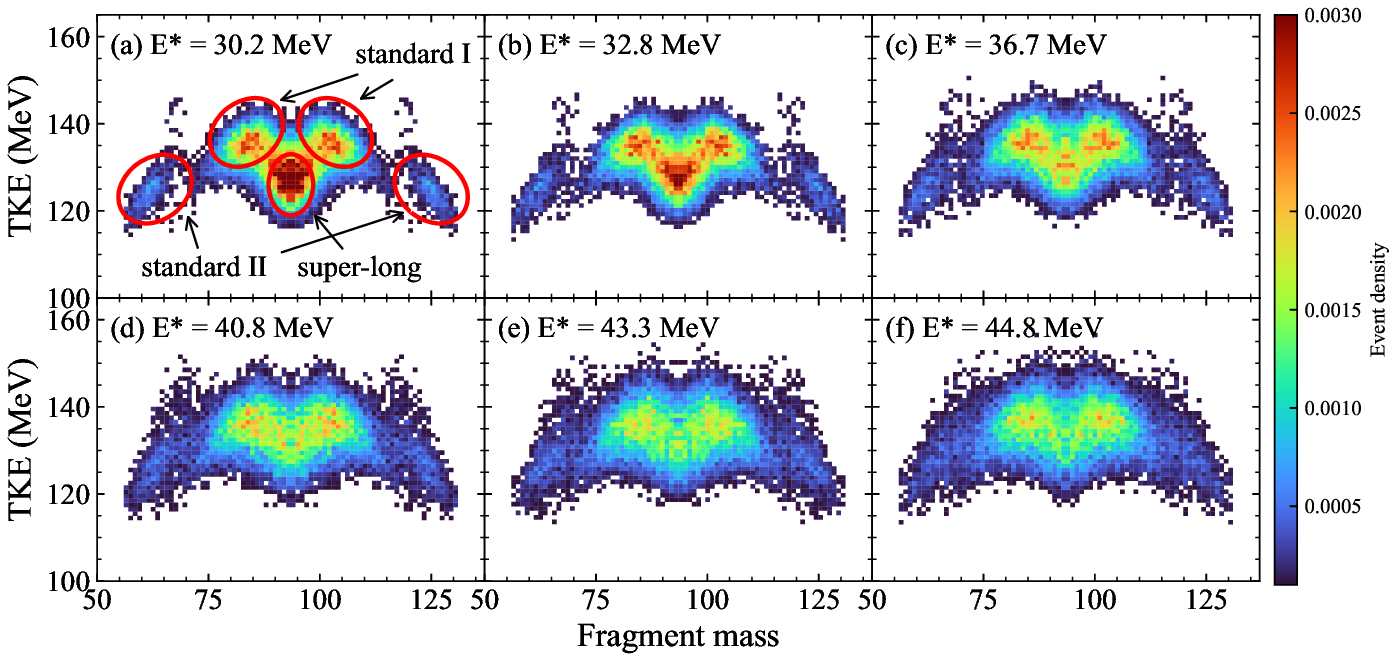}
\caption{\label{fig:mass_TKE_2d_all}The fragment mass-TKE correlations at six excitation energies.
Each subfigure is normalized to 1.}
\end{figure*}

\begin{figure}
\includegraphics[width=8.6cm]{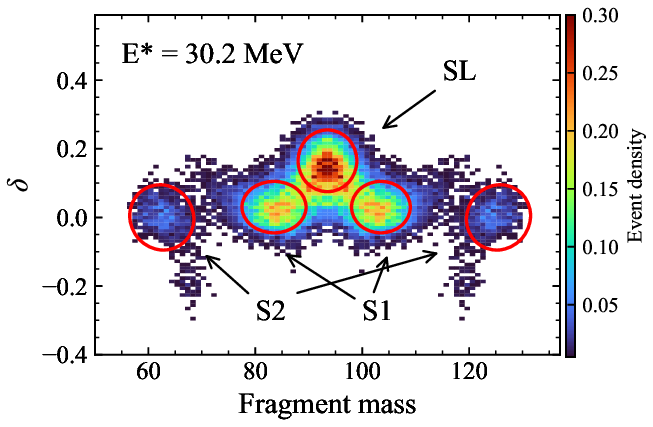}
\caption{\label{fig:mass_delta_2d_1E}The fragment deformation-mass ($\delta$-mass) correlations at $E^*=30.2$ MeV.
The distribution is normalized to 1.}
\end{figure}

The three-dimensional Langevin approach is utilized to calculate the mass, total kinetic energy (TKE), and deformation ($\delta$) of the $^{187}$Ir fission fragments at excitation energies $E^*=30.2$, 32.8, 36.7, 40.8, 43.3, and 44.8 MeV.
The fission barrier of $^{187}$Ir is 20 MeV.
To identify specific fission modes in $^{187}$Ir fission, the TKE-mass correlation is shown  in Fig.~\ref{fig:mass_TKE_2d_all}. 
Focusing first on shell-effect-led modes within the fragment mass range of 70 to 117, at low excitation energies, two distinct modes are apparent. 
One mode is symmetric, and another is an asymmetric fission mode at $A_L$/$A_H$ = 84/103.
The symmetric fission mode exhibits a lower fragment TKE, indicating an oblate deformation with a larger distance of the charge center, corresponding to $\delta$ = 0.15 as illustrated in Fig.~\ref{fig:mass_delta_2d_1E}.
This mode is termed the super-long (SL) mode. 
The asymmetric mode corresponds to a deformation of $\delta$ = 0, indicating a spherical shape, and is referred to as the standard I (S1) mode. 
The remaining asymmetric mode located at $A_L$/$A_H$ = 63/124 is denoted as the standard II (S2) mode.
In reference to actinide fission, the liquid drop (LD) behavior is also considered as a possible symmetric fission mode. 
To simplify naming, we refer to the LD behavior simply as the LD mode, deviating from the terminology used in actinide cases.
The general ranges of the SL, S1, and S2 are marked by ellipses in the mass-TKE correlation in Fig.~\ref{fig:mass_TKE_2d_all}(a).
As the excitation energy increases, the shell effects diminish, and the Langevin trajectories could explore to a wider range.
It leads to a more diffuse mass-TKE distributions at high excitation energies.

\begin{figure}
\includegraphics[width=8.6cm]{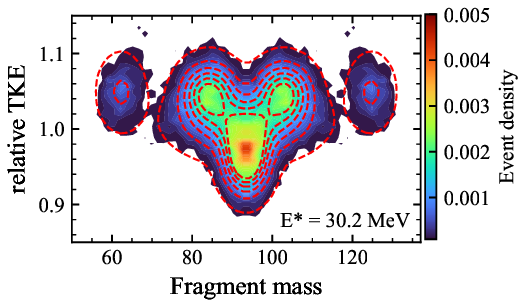}
\caption{\label{fig:mass_RTKE_2d_1E}The correlations of mass and relative TKE at $E^*=30.2$ MeV with the two-dimensional six-Gaussian fit shown by the red contour dashed lines.
The distribution is normalized to 1.}
\end{figure}

\begin{figure*}
\includegraphics[width=16.5cm]{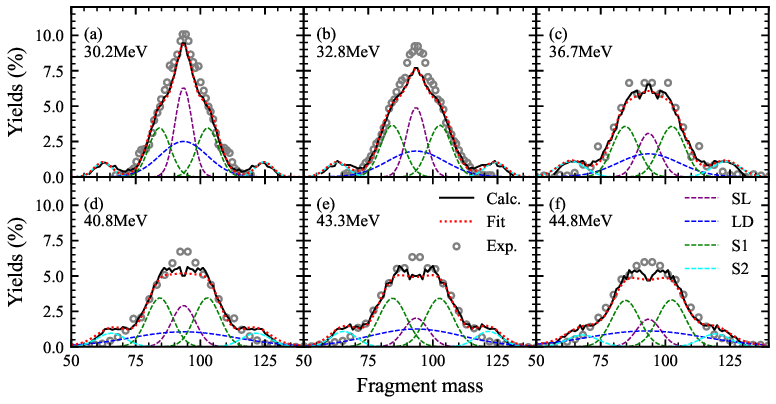}
\caption{\label{fig:mass_TKE_1d}The fission fragments mass distributions of $^{187}$Ir at six excitation energies with Gaussian fits of four fission modes.
The experimental data are taken from Ref.~\cite{PhysRevC.106.014616}.}
\end{figure*}

\begin{figure}
\includegraphics[width=8.6cm]{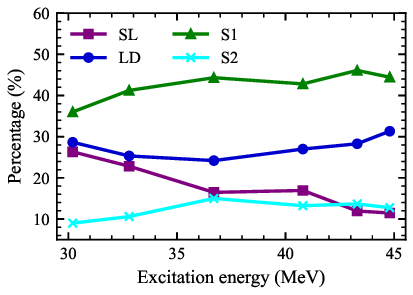}
\caption{\label{fig:mode_percent}The proportions of different modes as a function of excitation energies.
}
\end{figure}

To assess the proportions of each mode in mass distributions and their variation with excitation energy, we employ a two-dimensional fitting approach outlined in Ref.~\cite{SWINTONBLAND2023137655}. This involves a six-Gaussian fitting of the computed fragment mass and TKE.
To eliminate the dependence of TKE on mass, the relative TKE is used in fitting, calculated as the TKE divided by the Viola systematics value \cite{PhysRevC.31.1550,TOKE1985327}. Figure~\ref{fig:mass_RTKE_2d_1E} presents one of the two-dimensional fitting results. 
As it considers both fragment mass and TKE information, this approach provides enhanced precision compared to its one-dimensional counterpart, where only the mass distribution of the fragments is considered.
This heightened precision is particularly crucial as the fission of $^{187}$Ir involves two symmetric fission modes (the LD and the SL). 
Theirs proportions are challenging to determine through one-dimensional data.

Figure~\ref{fig:mass_TKE_1d} shows the calculated results of mass yields and the Gaussian fitting.
The dominant feature is a single-peak shape, 
which has well agreement with experimental data \cite{PhysRevC.106.014616},
while previous theoretical predictions give the double-peak shape \cite{PhysRevC.93.034620,PhysRevC.91.044316,SCHMIDT2016107}. 
The mass asymmetry of S2 is overestimated in our calculations.
The humps observed at $A_L$/$A_H$ = 77/110 in the experimental data might suggest the correct mass asymmetry of the S2 mode.
Our calculations reproduce the sudden change in slope and the sharp configuration of the main peak at 30.2 and 32.8 MeV. 
These two features indicate the superposition effects of the SL and S1 modes.
As the excitation energy increases, both calculations and experimental data exhibit an increase in the width and a decrease of peak height of the mass distribution.
This phenomenon arises from the attenuation of the shell effect at high excitation energies, which is estimated by a widely used phenomenological expression Eq.~(\ref{eq:phi}), and from a wider exploring allowed-ness of Langevin trajectories in the general coordinates space.
Although the overall dependence on excitation energy is reproduced, the results at 30.2 MeV and 40.8 MeV overestimate the excitation energy dependence of the symmetric peak and underestimate the energy dependence of the small asymmetric peaks on both sides.


In Fig.~\ref{fig:mode_percent}, the proportions of the four modes are depicted as a function of excitation energy. 
Notably, S1 and S2 exhibit higher resistance to excitation energy compared to SL, a feature consistent with observations in previous studies of sub-lead fission \cite{NISHIO201589, PRASAD2020135941, TSEKHANOVICH2019583, PhysRevC.104.024623, PhysRevC.105.014607, PhysRevC.107.034614, PhysRevC.108.054608}.
Within the Langevin approach framework, this resistance may be attributed to differences in the inertia tensor at different degrees of freedom. 
In the region from the saddle point to the scission point, the calculated inertia tensor element $m_{\eta,\eta}$ is about one order of magnitude higher than $m_{\delta,\delta}$. 
This suggests that, for prefragments, deformation is more easily accommodated than the exchange of nucleons under the same potential energy gradient. 
Consequently, the SL mode with an oblate shape becomes more susceptible to transitioning to other spherical modes as excitation energy increases, while the spherical shape of the asymmetric modes S1 and S2 imparts excitation resistance.
The change in the SL proportion elucidates the observed variation in the peak structure of the mass distribution. 
Despite the presence of the asymmetric fission mode S1 in $^{187}$Ir fission, the coexistence of the symmetric fission mode SL prevents the mass distribution from exhibiting a double-peak shape. 
This may be an aspect overlooked by other theoretical approaches, leading to double-peak fragment mass distributions in their calculations.

\begin{figure}
\includegraphics[width=8.6cm]{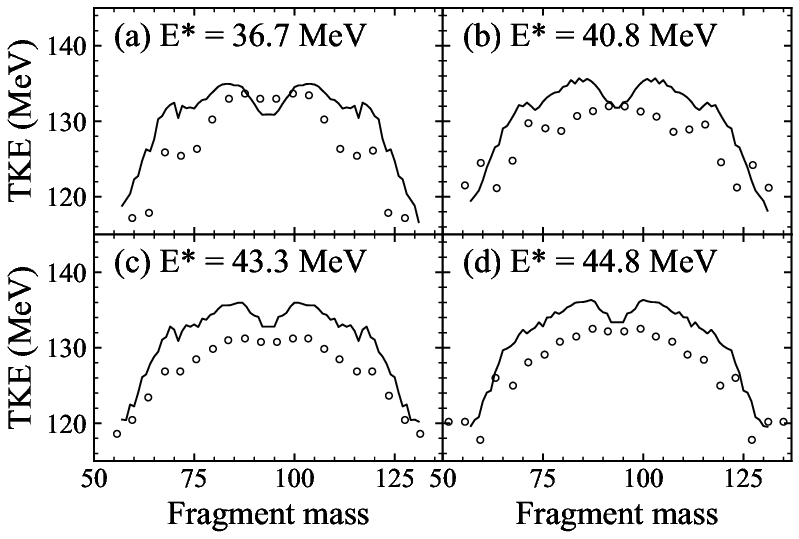}
\caption{\label{fig:mass_TKE_1d_4E}The calculated (solid line) and experimental \cite{PhysRevC.106.014616} (open circle) $\left< \mathrm{TKE} \right>$ of fission fragments as a function of mass.
}
\end{figure}

Table~\ref{tab:average_TKE} provides a comparison between the experimental and calculated $\left< \mathrm{TKE} \right>$ of fragments, which are approximately 130 MeV. 
In Fig.~\ref{fig:mass_TKE_1d_4E}, the experimental and calculated fragment TKE as a function of mass are compared. 
There is a deviation in $\left< \mathrm{TKE} \right>$ calculation, resulting in a global shift in the calculated curve compared to the experimental values.
In accordance with the liquid drop model, fission fragments exhibit a parabolic mass-TKE correlation \cite{TOKE1985327}. 
Any deviation from this parabolic trend suggests the potential presence of shell effects. 
It is observed that both the calculations and experimental data show a small peak around $A_L$ = 70 due to the effects of S2 mode. 
At the position of the symmetric fission, the calculated curve displays an obvious concavity attributed to the presence of the SL mode. 
A similar concavity can be observed in the experiment, albeit with a smaller magnitude.

\begin{table}[]
\caption{\label{tab:average_TKE}Experimental and calculated $\left< \mathrm{TKE} \right>$ of fragments at different excitation energy.}
\begin{ruledtabular}
\begin{tabular}{@{}ccc@{}}
\begin{tabular}[c]{@{}c@{}}Excitation\\ energy {(}MeV{)}\end{tabular} & Exp.\footnote{Experimental data are taken from \cite{PhysRevC.106.014616}}  & \begin{tabular}[c]{@{}c@{}}Langevin\end{tabular} \\
\colrule
30.2             & - & 130.5 \\
32.8             & - & 131.3 \\
36.7             & 130.9$\pm$0.5 & 132.3 \\
40.8             & 130.1$\pm$0.3 & 133.0 \\
43.3             & 129.6$\pm$0.4 & 133.5 \\
44.8             & 130.7$\pm$0.4 & 133.8 \\ 
\end{tabular}
\end{ruledtabular}
\end{table}

\begin{figure}
\includegraphics[width=8.6cm]{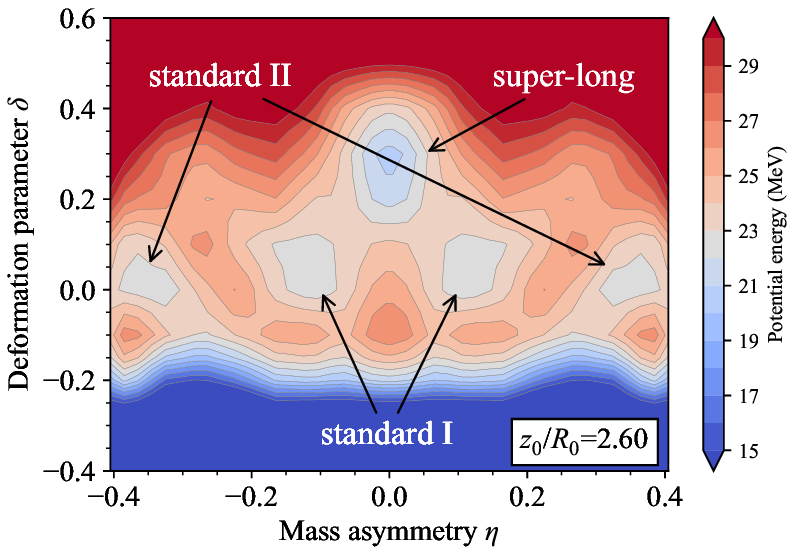}
\caption{\label{fig:V_eta_delta_2d}Potential energy surface at $z_0/R_0=2.6$.
}
\end{figure}

\begin{figure*}
\includegraphics[width=16.5cm]{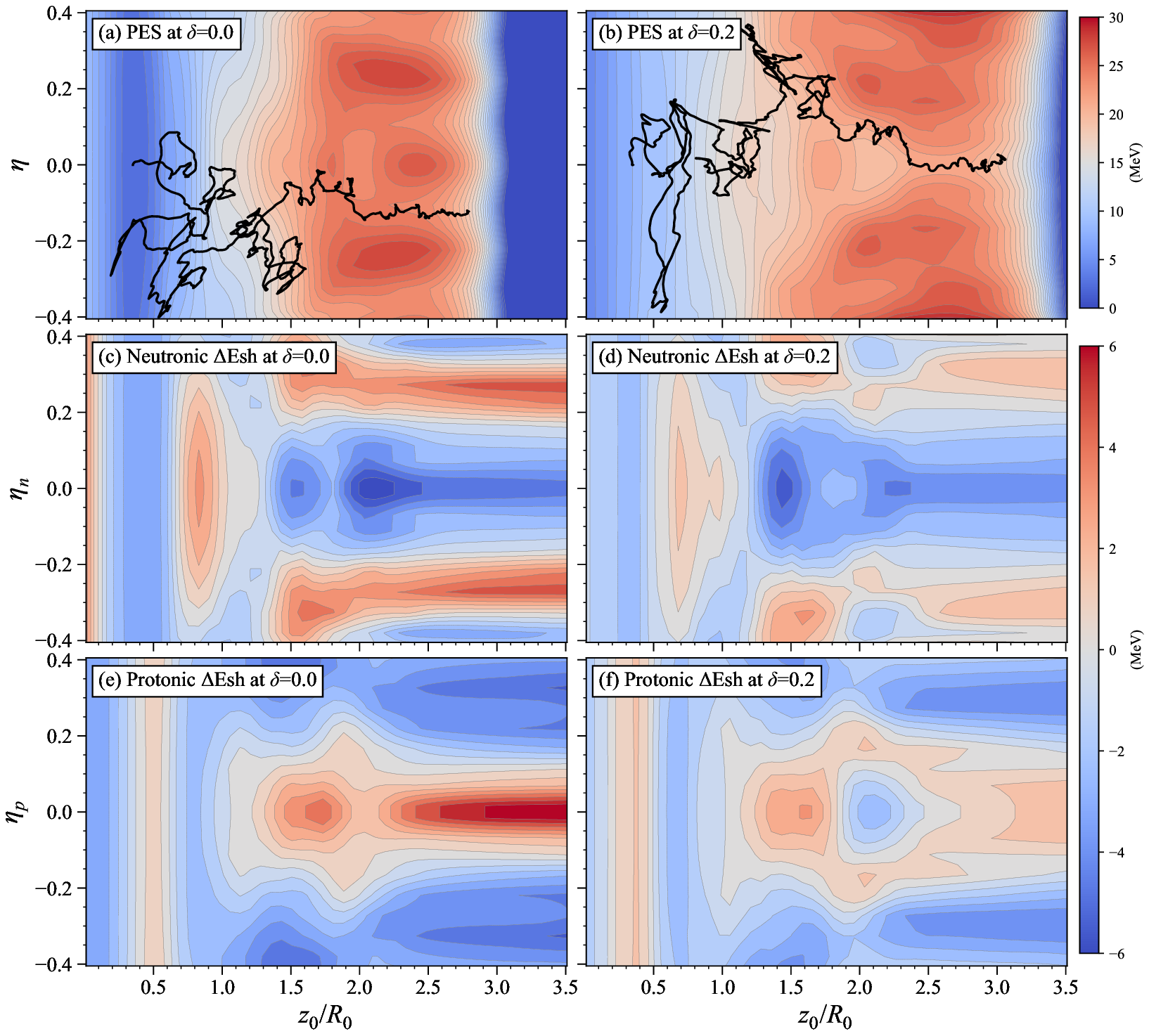}
\caption{\label{fig:trajectories_2d}(a),(b) The fission potential energy surfaces of $^{187}$Ir at $\delta=0$ and $0.2$. The black lines show the two Langevin trajectories.
(c),(d) The neutron shell correction energy surfaces at $\delta=0$ and $0.2$.
(e),(f) The proton shell correction energy surfaces at $\delta=0$ and $0.2$.}
\end{figure*}

To understand the origins of each fission mode, we examine the potential energy surface (PES) from various perspectives. 
Figure~\ref{fig:V_eta_delta_2d} illustrates the PES at $z_0$/$R_0$ = 2.6, revealing the potential energy channels associated with the three modes. 
It is observed that the $\delta$ of SL channel is 0.3, which is comparatively larger than the $\delta$ of the corresponding fragments in Fig~\ref{fig:mass_delta_2d_1E}. 
This is attributed to the higher elongation of the SL fragments. 
At larger elongation, the $\delta$ of the potential channel associated with SL becomes smaller.

Figure~\ref{fig:trajectories_2d} displays the PESs at two crucial nuclear deformation parameters, $\delta=0.0$ and 0.2.
Two Langevin fission trajectories are depicted as black lines.
The neutron and proton shell correction energy surfaces are also provided to elucidate the competition between neutron and proton shell effects. 
Assuming an unchanged charge distribution, the asymmetry parameters are identical for neutron, proton, and mass, i.e., $\eta=\eta_p=\eta_n$.
In Fig.~\ref{fig:trajectories_2d}(a), the asymmetry property of the S1 mode first emerges from a potential hill obstruction on the symmetric fission track near the elongation $z_0/R_0 = 1.2$. Although there is a small local minimum after the first hill, a second peak near the scission point impedes the trajectory toward symmetrical fission. 
It is noted that the PES does not show a long descent from saddle point to scission, which is consistent with the microscopic theoretical calculation \cite{PhysRevC.100.041602}. 
Additionally, the local minimum near $\left\{\eta=0.0, z_0/R_0=2.0\right\}$ suggests a possible fission isomer of $^{187}$Ir.
In Fig.~\ref{fig:trajectories_2d}(c), the neutron shell correction energy valley occurs at $\eta=0$ after $z_0/R_0=1.5$, while the proton correction energy valley occurs at $Z=50$ ($\eta_p=0.3$), as shown in Fig.~\ref{fig:trajectories_2d}(e). 
Noticeably, the peaks and valleys of the neutron and proton are almost complementary. 
Their superposition forms a new asymmetric potential energy valley located at the middle of the original peaks and valleys, around $\eta$ = 0.1. 
This superposition makes the PES calculation highly sensitive to the uncertainties of the neutron and proton shell correction energies. 
Such sensitivity is also evident in systematic calculations of $^{187}$Ir fission using the improved scission-point model \cite{PhysRevC.93.034620}. 
It may lead to the deviation in the predicted position of the potential valley around $\eta$ = 0.3, resulting in the appearance of humps in the mass distribution near $A_L$ = 63 (associated with S2 mode), rather than the experimentally observed location near $A_L$ = 77.

For the case of $\delta$ = 0.2, as shown in Fig.~\ref{fig:trajectories_2d}(b), a hill occurs near $z_0$/$R_0$ = 1.2. 
However, after bypassing the hill, the potential energy cliffs on both sides gradually tighten, forming a narrow potential energy valley. 
The trajectory is then confined around $\eta=0$. 
This restriction is more stringent than the one of LD behavior, leading to a narrower peak of mass yields at $E^*$ = 30.2 and 32.8 MeV. 
This narrow symmetric valley is mainly generated by a weaker deformed proton shell effect around $\eta_p=0.0$, as shown in Fig.~\ref{fig:trajectories_2d}(f), resulting in a lower symmetric shell correction valley than the one at $\delta$ = 0. 
A symmetric potential energy valley is then formed in the PES at $\delta$ = 0.2.

In general, unlike the actinide fission valley generated by the proton ($Z$ = 50) and neutron ($N$ = 82 or $N$ = 88) shell effects with close $\eta$, the valley of sub-lead fission would not necessarily lie at the same $\eta$ as that of the proton or neutron. 
This complexity increases the difficulty of identifying fission modes, especially based solely on magic numbers. 
For sub-lead nuclide fission, attention should be paid not only to whether the fragment contains magic numbers but also to the corresponding asymmetry parameter $\eta$. 
If the $\eta$ of neutron or proton shell effects has a gap, it is necessary to consider whether the position of the asymmetric mass yields peak would be shifted due to the superposition of shell effects.


\section{\label{sec:summary}Summary}
The calculated fission yields of $^{187}$Ir from various theoretical models significantly differ from experimental data, and the fission mode is not clear. 
This study employs the three-dimensional Langevin dynamics approach to calculate the mass, total kinetic energy (TKE), and deformation ($\delta$) of $^{187}$Ir fission fragments at different excitation energies. 
Four fission modes are identified based on mass-TKE and mass-$\delta$ correlations, namely two asymmetric standard modes S1 and S2, a symmetric super-long mode SL, and a symmetric liquid-drop mode LD.
The center of the asymmetric mass yield peak for S1 is located around $A_L$/$A_H$ = 84/103, and for S2, it is around $A_L$/$A_H$ = 63/124. 
The calculations reveal strong excitation-energy resistance of the asymmetric modes, a feature consistent with other sub-lead isotope fission. 
This resistance might be attributed to the relatively high inertia of collective motion caused by nucleon transfer between prefragments. 
The single-peak shape dominates the mass distribution, particularly at low excitation energies, aligning well with experimental data. 
The underestimation of the SL mode may explain previous predictions of double-peaked fragment mass distributions by other theoretical models.

Analysis of the fission potential energy surface indicates that the asymmetric potential energy valleys leading to S1 and S2 modes are not identical to any individual proton or neutron shell correction valley, differing from actinide fission. 
This complexity complicates the identification of fission modes, with the uncertainties of shell correction may play a crucial role. 
These results reveal a complex multimodality of $^{187}$Ir fission under its single-peak mass distribution, emphasizing the need to consider fission mode displacement when there is a gap in the asymmetry parameters ($\eta$) of the proton and neutron magic numbers possessed by fragments.


\section*{ACKNOWLEDGMENTS}

This work was supported by the Key Laboratory of Nuclear Data foundation No. JCKY2022201C157, and the National Natural Science Foundation of China under Grant No. 11875328.

%


\bibliography{apssamp_prc}

\begin{thebibliography}{52}%
\makeatletter
\providecommand \@ifxundefined [1]{%
 \@ifx{#1\undefined}
}%
\providecommand \@ifnum [1]{%
 \ifnum #1\expandafter \@firstoftwo
 \else \expandafter \@secondoftwo
 \fi
}%
\providecommand \@ifx [1]{%
 \ifx #1\expandafter \@firstoftwo
 \else \expandafter \@secondoftwo
 \fi
}%
\providecommand \natexlab [1]{#1}%
\providecommand \enquote  [1]{``#1''}%
\providecommand \bibnamefont  [1]{#1}%
\providecommand \bibfnamefont [1]{#1}%
\providecommand \citenamefont [1]{#1}%
\providecommand \href@noop [0]{\@secondoftwo}%
\providecommand \href [0]{\begingroup \@sanitize@url \@href}%
\providecommand \@href[1]{\@@startlink{#1}\@@href}%
\providecommand \@@href[1]{\endgroup#1\@@endlink}%
\providecommand \@sanitize@url [0]{\catcode `\\12\catcode `\$12\catcode
  `\&12\catcode `\#12\catcode `\^12\catcode `\_12\catcode `\%12\relax}%
\providecommand \@@startlink[1]{}%
\providecommand \@@endlink[0]{}%
\providecommand \url  [0]{\begingroup\@sanitize@url \@url }%
\providecommand \@url [1]{\endgroup\@href {#1}{\urlprefix }}%
\providecommand \urlprefix  [0]{URL }%
\providecommand \Eprint [0]{\href }%
\providecommand \doibase [0]{http://dx.doi.org/}%
\providecommand \selectlanguage [0]{\@gobble}%
\providecommand \bibinfo  [0]{\@secondoftwo}%
\providecommand \bibfield  [0]{\@secondoftwo}%
\providecommand \translation [1]{[#1]}%
\providecommand \BibitemOpen [0]{}%
\providecommand \bibitemStop [0]{}%
\providecommand \bibitemNoStop [0]{.\EOS\space}%
\providecommand \EOS [0]{\spacefactor3000\relax}%
\providecommand \BibitemShut  [1]{\csname bibitem#1\endcsname}%
\let\auto@bib@innerbib\@empty
\bibitem [{\citenamefont {Meitner}\ and\ \citenamefont
  {Frisch}(1939)}]{Meitner1939239}%
  \BibitemOpen
  \bibfield  {author} {\bibinfo {author} {\bibfnamefont {L.}~\bibnamefont
  {Meitner}}\ and\ \bibinfo {author} {\bibfnamefont {O.}~\bibnamefont
  {Frisch}},\ }\href {\doibase 10.1038/143239a0} {\bibfield  {journal}
  {\bibinfo  {journal} {Nature}\ }\textbf {\bibinfo {volume} {143}},\ \bibinfo
  {pages} {239 – 240} (\bibinfo {year} {1939})}\BibitemShut {NoStop}%
\bibitem [{\citenamefont {Bohr}\ and\ \citenamefont
  {Wheeler}(1939)}]{Bohr1939426}%
  \BibitemOpen
  \bibfield  {author} {\bibinfo {author} {\bibfnamefont {N.}~\bibnamefont
  {Bohr}}\ and\ \bibinfo {author} {\bibfnamefont {J.~A.}\ \bibnamefont
  {Wheeler}},\ }\href {\doibase 10.1103/PhysRev.56.426} {\bibfield  {journal}
  {\bibinfo  {journal} {Phys. Rev.}\ }\textbf {\bibinfo {volume} {56}},\
  \bibinfo {pages} {426 – 450} (\bibinfo {year} {1939})}\BibitemShut
  {NoStop}%
\bibitem [{\citenamefont {Strutinsky}(1967)}]{Strutinsky-98}%
  \BibitemOpen
  \bibfield  {author} {\bibinfo {author} {\bibfnamefont {V.~M.}\ \bibnamefont
  {Strutinsky}},\ }\href {\doibase
  https://doi.org/10.1016/0375-9474(67)90510-6} {\bibfield  {journal} {\bibinfo
   {journal} {Nucl. Phys. A}\ }\textbf {\bibinfo {volume} {95}},\ \bibinfo
  {pages} {420} (\bibinfo {year} {1967})}\BibitemShut {NoStop}%
\bibitem [{\citenamefont {Strutinsky}(1968)}]{STRUTINSKY19681}%
  \BibitemOpen
  \bibfield  {author} {\bibinfo {author} {\bibfnamefont {V.~M.}\ \bibnamefont
  {Strutinsky}},\ }\href {\doibase
  https://doi.org/10.1016/0375-9474(68)90699-4} {\bibfield  {journal} {\bibinfo
   {journal} {Nucl. Phys. A}\ }\textbf {\bibinfo {volume} {122}},\ \bibinfo
  {pages} {1} (\bibinfo {year} {1968})}\BibitemShut {NoStop}%
\bibitem [{\citenamefont {Brosa}\ \emph {et~al.}(1990)\citenamefont {Brosa},
  \citenamefont {Grossmann},\ and\ \citenamefont {Müller}}]{BROSA1990167}%
  \BibitemOpen
  \bibfield  {author} {\bibinfo {author} {\bibfnamefont {U.}~\bibnamefont
  {Brosa}}, \bibinfo {author} {\bibfnamefont {S.}~\bibnamefont {Grossmann}}, \
  and\ \bibinfo {author} {\bibfnamefont {A.}~\bibnamefont {Müller}},\ }\href
  {\doibase https://doi.org/10.1016/0370-1573(90)90114-H} {\bibfield  {journal}
  {\bibinfo  {journal} {Phys. Rep.}\ }\textbf {\bibinfo {volume} {197}},\
  \bibinfo {pages} {167} (\bibinfo {year} {1990})}\BibitemShut {NoStop}%
\bibitem [{\citenamefont {Schmidt}\ \emph {et~al.}(2000)\citenamefont
  {Schmidt}, \citenamefont {Steinhäuser}, \citenamefont {Böckstiegel},
  \citenamefont {Grewe}, \citenamefont {Heinz}, \citenamefont {Junghans},
  \citenamefont {Benlliure}, \citenamefont {Clerc}, \citenamefont {{de Jong}},
  \citenamefont {Müller} \emph {et~al.}}]{SCHMIDT2000221}%
  \BibitemOpen
  \bibfield  {author} {\bibinfo {author} {\bibfnamefont {K.-H.}\ \bibnamefont
  {Schmidt}}, \bibinfo {author} {\bibfnamefont {S.}~\bibnamefont
  {Steinhäuser}}, \bibinfo {author} {\bibfnamefont {C.}~\bibnamefont
  {Böckstiegel}}, \bibinfo {author} {\bibfnamefont {A.}~\bibnamefont {Grewe}},
  \bibinfo {author} {\bibfnamefont {A.}~\bibnamefont {Heinz}}, \bibinfo
  {author} {\bibfnamefont {A.}~\bibnamefont {Junghans}}, \bibinfo {author}
  {\bibfnamefont {J.}~\bibnamefont {Benlliure}}, \bibinfo {author}
  {\bibfnamefont {H.-G.}\ \bibnamefont {Clerc}}, \bibinfo {author}
  {\bibfnamefont {M.}~\bibnamefont {{de Jong}}}, \bibinfo {author}
  {\bibfnamefont {J.}~\bibnamefont {Müller}},  \emph {et~al.},\ }\href
  {\doibase https://doi.org/10.1016/S0375-9474(99)00384-X} {\bibfield
  {journal} {\bibinfo  {journal} {Nucl. Phys. A}\ }\textbf {\bibinfo {volume}
  {665}},\ \bibinfo {pages} {221} (\bibinfo {year} {2000})}\BibitemShut
  {NoStop}%
\bibitem [{\citenamefont {Scamps}\ and\ \citenamefont
  {Simenel}(2018)}]{scamps2018impact}%
  \BibitemOpen
  \bibfield  {author} {\bibinfo {author} {\bibfnamefont {G.}~\bibnamefont
  {Scamps}}\ and\ \bibinfo {author} {\bibfnamefont {C.}~\bibnamefont
  {Simenel}},\ }\href {\doibase https://doi.org/10.1038/s41586-018-0780-0}
  {\bibfield  {journal} {\bibinfo  {journal} {Nature}\ }\textbf {\bibinfo
  {volume} {564}},\ \bibinfo {pages} {382} (\bibinfo {year}
  {2018})}\BibitemShut {NoStop}%
\bibitem [{\citenamefont {Mahata}\ \emph {et~al.}(2022)\citenamefont {Mahata},
  \citenamefont {Schmitt}, \citenamefont {Gupta}, \citenamefont {Shrivastava},
  \citenamefont {Scamps},\ and\ \citenamefont {Schmidt}}]{MAHATA2022136859}%
  \BibitemOpen
  \bibfield  {author} {\bibinfo {author} {\bibfnamefont {K.}~\bibnamefont
  {Mahata}}, \bibinfo {author} {\bibfnamefont {C.}~\bibnamefont {Schmitt}},
  \bibinfo {author} {\bibfnamefont {S.}~\bibnamefont {Gupta}}, \bibinfo
  {author} {\bibfnamefont {A.}~\bibnamefont {Shrivastava}}, \bibinfo {author}
  {\bibfnamefont {G.}~\bibnamefont {Scamps}}, \ and\ \bibinfo {author}
  {\bibfnamefont {K.-H.}\ \bibnamefont {Schmidt}},\ }\href {\doibase
  https://doi.org/10.1016/j.physletb.2021.136859} {\bibfield  {journal}
  {\bibinfo  {journal} {Phys. Lett. B}\ }\textbf {\bibinfo {volume} {825}},\
  \bibinfo {pages} {136859} (\bibinfo {year} {2022})}\BibitemShut {NoStop}%
\bibitem [{\citenamefont {Qiao}\ \emph {et~al.}(2021)\citenamefont {Qiao},
  \citenamefont {Pei}, \citenamefont {Wang}, \citenamefont {Qiang},
  \citenamefont {Chen}, \citenamefont {Shu},\ and\ \citenamefont
  {Ge}}]{PhysRevC.103.034621}%
  \BibitemOpen
  \bibfield  {author} {\bibinfo {author} {\bibfnamefont {C.~Y.}\ \bibnamefont
  {Qiao}}, \bibinfo {author} {\bibfnamefont {J.~C.}\ \bibnamefont {Pei}},
  \bibinfo {author} {\bibfnamefont {Z.~A.}\ \bibnamefont {Wang}}, \bibinfo
  {author} {\bibfnamefont {Y.}~\bibnamefont {Qiang}}, \bibinfo {author}
  {\bibfnamefont {Y.~J.}\ \bibnamefont {Chen}}, \bibinfo {author}
  {\bibfnamefont {N.~C.}\ \bibnamefont {Shu}}, \ and\ \bibinfo {author}
  {\bibfnamefont {Z.~G.}\ \bibnamefont {Ge}},\ }\href {\doibase
  10.1103/PhysRevC.103.034621} {\bibfield  {journal} {\bibinfo  {journal}
  {Phys. Rev. C}\ }\textbf {\bibinfo {volume} {103}},\ \bibinfo {pages}
  {034621} (\bibinfo {year} {2021})}\BibitemShut {NoStop}%
\bibitem [{int(2008)}]{international2008iaea}%
  \BibitemOpen
  \href
  {https://www.iaea.org/publications/7659/fission-product-yield-data-for-the-transmutation-of-minor-actinide-nuclear-waste}
  {\emph {\bibinfo {title} {Fission Product Yield Data for the Transmutation of
  Minor Actinide Nuclear Waste}}},\ Non-serial Publications\ (\bibinfo
  {publisher} {International Atomic Energy Agency},\ \bibinfo {address}
  {Vienna},\ \bibinfo {year} {2008})\BibitemShut {NoStop}%
\bibitem [{\citenamefont {Andreyev}\ \emph {et~al.}(2010)\citenamefont
  {Andreyev}, \citenamefont {Elseviers}, \citenamefont {Huyse}, \citenamefont
  {Van~Duppen}, \citenamefont {Antalic}, \citenamefont {Barzakh}, \citenamefont
  {Bree}, \citenamefont {Cocolios}, \citenamefont {Comas}, \citenamefont
  {Diriken} \emph {et~al.}}]{PhysRevLett.105.252502}%
  \BibitemOpen
  \bibfield  {author} {\bibinfo {author} {\bibfnamefont {A.~N.}\ \bibnamefont
  {Andreyev}}, \bibinfo {author} {\bibfnamefont {J.}~\bibnamefont {Elseviers}},
  \bibinfo {author} {\bibfnamefont {M.}~\bibnamefont {Huyse}}, \bibinfo
  {author} {\bibfnamefont {P.}~\bibnamefont {Van~Duppen}}, \bibinfo {author}
  {\bibfnamefont {S.}~\bibnamefont {Antalic}}, \bibinfo {author} {\bibfnamefont
  {A.}~\bibnamefont {Barzakh}}, \bibinfo {author} {\bibfnamefont
  {N.}~\bibnamefont {Bree}}, \bibinfo {author} {\bibfnamefont {T.~E.}\
  \bibnamefont {Cocolios}}, \bibinfo {author} {\bibfnamefont {V.~F.}\
  \bibnamefont {Comas}}, \bibinfo {author} {\bibfnamefont {J.}~\bibnamefont
  {Diriken}},  \emph {et~al.},\ }\href {\doibase
  10.1103/PhysRevLett.105.252502} {\bibfield  {journal} {\bibinfo  {journal}
  {Phys. Rev. Lett.}\ }\textbf {\bibinfo {volume} {105}},\ \bibinfo {pages}
  {252502} (\bibinfo {year} {2010})}\BibitemShut {NoStop}%
\bibitem [{\citenamefont {M\"oller}\ \emph {et~al.}(2012)\citenamefont
  {M\"oller}, \citenamefont {Randrup},\ and\ \citenamefont
  {Sierk}}]{PhysRevC.85.024306}%
  \BibitemOpen
  \bibfield  {author} {\bibinfo {author} {\bibfnamefont {P.}~\bibnamefont
  {M\"oller}}, \bibinfo {author} {\bibfnamefont {J.}~\bibnamefont {Randrup}}, \
  and\ \bibinfo {author} {\bibfnamefont {A.~J.}\ \bibnamefont {Sierk}},\ }\href
  {\doibase 10.1103/PhysRevC.85.024306} {\bibfield  {journal} {\bibinfo
  {journal} {Phys. Rev. C}\ }\textbf {\bibinfo {volume} {85}},\ \bibinfo
  {pages} {024306} (\bibinfo {year} {2012})}\BibitemShut {NoStop}%
\bibitem [{\citenamefont {McDonnell}\ \emph {et~al.}(2014)\citenamefont
  {McDonnell}, \citenamefont {Nazarewicz}, \citenamefont {Sheikh},
  \citenamefont {Staszczak},\ and\ \citenamefont {Warda}}]{PhysRevC.90.021302}%
  \BibitemOpen
  \bibfield  {author} {\bibinfo {author} {\bibfnamefont {J.~D.}\ \bibnamefont
  {McDonnell}}, \bibinfo {author} {\bibfnamefont {W.}~\bibnamefont
  {Nazarewicz}}, \bibinfo {author} {\bibfnamefont {J.~A.}\ \bibnamefont
  {Sheikh}}, \bibinfo {author} {\bibfnamefont {A.}~\bibnamefont {Staszczak}}, \
  and\ \bibinfo {author} {\bibfnamefont {M.}~\bibnamefont {Warda}},\ }\href
  {\doibase 10.1103/PhysRevC.90.021302} {\bibfield  {journal} {\bibinfo
  {journal} {Phys. Rev. C}\ }\textbf {\bibinfo {volume} {90}},\ \bibinfo
  {pages} {021302(R)} (\bibinfo {year} {2014})}\BibitemShut {NoStop}%
\bibitem [{\citenamefont {Scamps}\ and\ \citenamefont
  {Simenel}(2019)}]{PhysRevC.100.041602}%
  \BibitemOpen
  \bibfield  {author} {\bibinfo {author} {\bibfnamefont {G.}~\bibnamefont
  {Scamps}}\ and\ \bibinfo {author} {\bibfnamefont {C.}~\bibnamefont
  {Simenel}},\ }\href {\doibase 10.1103/PhysRevC.100.041602} {\bibfield
  {journal} {\bibinfo  {journal} {Phys. Rev. C}\ }\textbf {\bibinfo {volume}
  {100}},\ \bibinfo {pages} {041602(R)} (\bibinfo {year} {2019})}\BibitemShut
  {NoStop}%
\bibitem [{\citenamefont {Ichikawa}\ and\ \citenamefont
  {M{\"o}ller}(2019)}]{ichikawa2019microscopic}%
  \BibitemOpen
  \bibfield  {author} {\bibinfo {author} {\bibfnamefont {T.}~\bibnamefont
  {Ichikawa}}\ and\ \bibinfo {author} {\bibfnamefont {P.}~\bibnamefont
  {M{\"o}ller}},\ }\href@noop {} {\bibfield  {journal} {\bibinfo  {journal}
  {Phys. Lett. B}\ }\textbf {\bibinfo {volume} {789}},\ \bibinfo {pages} {679}
  (\bibinfo {year} {2019})}\BibitemShut {NoStop}%
\bibitem [{\citenamefont {Li}\ \emph {et~al.}(2022)\citenamefont {Li},
  \citenamefont {Chen}, \citenamefont {Chen},\ and\ \citenamefont
  {Li}}]{PhysRevC.106.024307}%
  \BibitemOpen
  \bibfield  {author} {\bibinfo {author} {\bibfnamefont {Z.}~\bibnamefont
  {Li}}, \bibinfo {author} {\bibfnamefont {S.}~\bibnamefont {Chen}}, \bibinfo
  {author} {\bibfnamefont {Y.}~\bibnamefont {Chen}}, \ and\ \bibinfo {author}
  {\bibfnamefont {Z.}~\bibnamefont {Li}},\ }\href {\doibase
  10.1103/PhysRevC.106.024307} {\bibfield  {journal} {\bibinfo  {journal}
  {Phys. Rev. C}\ }\textbf {\bibinfo {volume} {106}},\ \bibinfo {pages}
  {024307} (\bibinfo {year} {2022})}\BibitemShut {NoStop}%
\bibitem [{\citenamefont {Bernard}\ \emph {et~al.}(2023)\citenamefont
  {Bernard}, \citenamefont {Simenel},\ and\ \citenamefont
  {Blanchon}}]{bernard2023hartree}%
  \BibitemOpen
  \bibfield  {author} {\bibinfo {author} {\bibfnamefont {R.}~\bibnamefont
  {Bernard}}, \bibinfo {author} {\bibfnamefont {C.}~\bibnamefont {Simenel}}, \
  and\ \bibinfo {author} {\bibfnamefont {G.}~\bibnamefont {Blanchon}},\
  }\href@noop {} {\bibfield  {journal} {\bibinfo  {journal} {Eur. Phys. J. A}\
  }\textbf {\bibinfo {volume} {59}},\ \bibinfo {pages} {51} (\bibinfo {year}
  {2023})}\BibitemShut {NoStop}%
\bibitem [{\citenamefont {Usang}\ \emph {et~al.}(2019)\citenamefont {Usang},
  \citenamefont {Ivanyuk}, \citenamefont {Ishizuka},\ and\ \citenamefont
  {Chiba}}]{usang2018}%
  \BibitemOpen
  \bibfield  {author} {\bibinfo {author} {\bibfnamefont {M.~D.}\ \bibnamefont
  {Usang}}, \bibinfo {author} {\bibfnamefont {F.~A.}\ \bibnamefont {Ivanyuk}},
  \bibinfo {author} {\bibfnamefont {C.}~\bibnamefont {Ishizuka}}, \ and\
  \bibinfo {author} {\bibfnamefont {S.}~\bibnamefont {Chiba}},\ }\href
  {\doibase https://doi.org/10.1038/s41598-018-37993-7} {\bibfield  {journal}
  {\bibinfo  {journal} {Sci. Rep.}\ }\textbf {\bibinfo {volume} {9}},\ \bibinfo
  {pages} {1525} (\bibinfo {year} {2019})}\BibitemShut {NoStop}%
\bibitem [{\citenamefont {Swinton-Bland}\ \emph {et~al.}(2023)\citenamefont
  {Swinton-Bland}, \citenamefont {Buete}, \citenamefont {Hinde}, \citenamefont
  {Dasgupta}, \citenamefont {Tanaka}, \citenamefont {Berriman}, \citenamefont
  {Jeung}, \citenamefont {Banerjee}, \citenamefont {Bezzina}, \citenamefont
  {Carter} \emph {et~al.}}]{SWINTONBLAND2023137655}%
  \BibitemOpen
  \bibfield  {author} {\bibinfo {author} {\bibfnamefont {B.}~\bibnamefont
  {Swinton-Bland}}, \bibinfo {author} {\bibfnamefont {J.}~\bibnamefont
  {Buete}}, \bibinfo {author} {\bibfnamefont {D.}~\bibnamefont {Hinde}},
  \bibinfo {author} {\bibfnamefont {M.}~\bibnamefont {Dasgupta}}, \bibinfo
  {author} {\bibfnamefont {T.}~\bibnamefont {Tanaka}}, \bibinfo {author}
  {\bibfnamefont {A.}~\bibnamefont {Berriman}}, \bibinfo {author}
  {\bibfnamefont {D.}~\bibnamefont {Jeung}}, \bibinfo {author} {\bibfnamefont
  {K.}~\bibnamefont {Banerjee}}, \bibinfo {author} {\bibfnamefont
  {L.}~\bibnamefont {Bezzina}}, \bibinfo {author} {\bibfnamefont
  {I.}~\bibnamefont {Carter}},  \emph {et~al.},\ }\href {\doibase
  https://doi.org/10.1016/j.physletb.2022.137655} {\bibfield  {journal}
  {\bibinfo  {journal} {Phys. Lett. B}\ }\textbf {\bibinfo {volume} {837}},\
  \bibinfo {pages} {137655} (\bibinfo {year} {2023})}\BibitemShut {NoStop}%
\bibitem [{\citenamefont {Andreev}\ \emph {et~al.}(2016)\citenamefont
  {Andreev}, \citenamefont {Adamian},\ and\ \citenamefont
  {Antonenko}}]{PhysRevC.93.034620}%
  \BibitemOpen
  \bibfield  {author} {\bibinfo {author} {\bibfnamefont {A.~V.}\ \bibnamefont
  {Andreev}}, \bibinfo {author} {\bibfnamefont {G.~G.}\ \bibnamefont
  {Adamian}}, \ and\ \bibinfo {author} {\bibfnamefont {N.~V.}\ \bibnamefont
  {Antonenko}},\ }\href {\doibase 10.1103/PhysRevC.93.034620} {\bibfield
  {journal} {\bibinfo  {journal} {Phys. Rev. C}\ }\textbf {\bibinfo {volume}
  {93}},\ \bibinfo {pages} {034620} (\bibinfo {year} {2016})}\BibitemShut
  {NoStop}%
\bibitem [{\citenamefont {M\"oller}\ and\ \citenamefont
  {Randrup}(2015)}]{PhysRevC.91.044316}%
  \BibitemOpen
  \bibfield  {author} {\bibinfo {author} {\bibfnamefont {P.}~\bibnamefont
  {M\"oller}}\ and\ \bibinfo {author} {\bibfnamefont {J.}~\bibnamefont
  {Randrup}},\ }\href {\doibase 10.1103/PhysRevC.91.044316} {\bibfield
  {journal} {\bibinfo  {journal} {Phys. Rev. C}\ }\textbf {\bibinfo {volume}
  {91}},\ \bibinfo {pages} {044316} (\bibinfo {year} {2015})}\BibitemShut
  {NoStop}%
\bibitem [{\citenamefont {Schmidt}\ \emph {et~al.}(2016)\citenamefont
  {Schmidt}, \citenamefont {Jurado}, \citenamefont {Amouroux},\ and\
  \citenamefont {Schmitt}}]{SCHMIDT2016107}%
  \BibitemOpen
  \bibfield  {author} {\bibinfo {author} {\bibfnamefont {K.-H.}\ \bibnamefont
  {Schmidt}}, \bibinfo {author} {\bibfnamefont {B.}~\bibnamefont {Jurado}},
  \bibinfo {author} {\bibfnamefont {C.}~\bibnamefont {Amouroux}}, \ and\
  \bibinfo {author} {\bibfnamefont {C.}~\bibnamefont {Schmitt}},\ }\href
  {\doibase https://doi.org/10.1016/j.nds.2015.12.009} {\bibfield  {journal}
  {\bibinfo  {journal} {Nucl. Data Sheets}\ }\textbf {\bibinfo {volume}
  {131}},\ \bibinfo {pages} {107} (\bibinfo {year} {2016})}\BibitemShut
  {NoStop}%
\bibitem [{\citenamefont {Itkis}\ \emph {et~al.}(1991)\citenamefont {Itkis},
  \citenamefont {Kondratev}, \citenamefont {Mulgin}, \citenamefont {Okolovich},
  \citenamefont {Rusanov},\ and\ \citenamefont {Smirenkin}}]{itkis1991}%
  \BibitemOpen
  \bibfield  {author} {\bibinfo {author} {\bibfnamefont {M.~G.}\ \bibnamefont
  {Itkis}}, \bibinfo {author} {\bibfnamefont {N.~A.}\ \bibnamefont
  {Kondratev}}, \bibinfo {author} {\bibfnamefont {S.~I.}\ \bibnamefont
  {Mulgin}}, \bibinfo {author} {\bibfnamefont {V.~N.}\ \bibnamefont
  {Okolovich}}, \bibinfo {author} {\bibfnamefont {A.~Y.}\ \bibnamefont
  {Rusanov}}, \ and\ \bibinfo {author} {\bibfnamefont {G.~N.}\ \bibnamefont
  {Smirenkin}},\ }\href@noop {} {\bibfield  {journal} {\bibinfo  {journal}
  {Yad. Fiz.}\ }\textbf {\bibinfo {volume} {53}},\ \bibinfo {pages} {1225}
  (\bibinfo {year} {1991})}\BibitemShut {NoStop}%
\bibitem [{\citenamefont {Dhuri}\ \emph {et~al.}(2022)\citenamefont {Dhuri},
  \citenamefont {Mahata}, \citenamefont {Shrivastava}, \citenamefont
  {Ramachandran}, \citenamefont {Pandit}, \citenamefont {Kumar}, \citenamefont
  {Parkar}, \citenamefont {Rout}, \citenamefont {Kumar}, \citenamefont {Chavan}
  \emph {et~al.}}]{PhysRevC.106.014616}%
  \BibitemOpen
  \bibfield  {author} {\bibinfo {author} {\bibfnamefont {S.}~\bibnamefont
  {Dhuri}}, \bibinfo {author} {\bibfnamefont {K.}~\bibnamefont {Mahata}},
  \bibinfo {author} {\bibfnamefont {A.}~\bibnamefont {Shrivastava}}, \bibinfo
  {author} {\bibfnamefont {K.}~\bibnamefont {Ramachandran}}, \bibinfo {author}
  {\bibfnamefont {S.~K.}\ \bibnamefont {Pandit}}, \bibinfo {author}
  {\bibfnamefont {V.}~\bibnamefont {Kumar}}, \bibinfo {author} {\bibfnamefont
  {V.~V.}\ \bibnamefont {Parkar}}, \bibinfo {author} {\bibfnamefont {P.~C.}\
  \bibnamefont {Rout}}, \bibinfo {author} {\bibfnamefont {A.}~\bibnamefont
  {Kumar}}, \bibinfo {author} {\bibfnamefont {A.}~\bibnamefont {Chavan}},
  \emph {et~al.},\ }\href {\doibase 10.1103/PhysRevC.106.014616} {\bibfield
  {journal} {\bibinfo  {journal} {Phys. Rev. C}\ }\textbf {\bibinfo {volume}
  {106}},\ \bibinfo {pages} {014616} (\bibinfo {year} {2022})}\BibitemShut
  {NoStop}%
\bibitem [{\citenamefont {Ishizuka}\ \emph {et~al.}(2017)\citenamefont
  {Ishizuka}, \citenamefont {Usang}, \citenamefont {Ivanyuk}, \citenamefont
  {Maruhn}, \citenamefont {Nishio},\ and\ \citenamefont
  {Chiba}}]{PhysRevC.96.064616}%
  \BibitemOpen
  \bibfield  {author} {\bibinfo {author} {\bibfnamefont {C.}~\bibnamefont
  {Ishizuka}}, \bibinfo {author} {\bibfnamefont {M.~D.}\ \bibnamefont {Usang}},
  \bibinfo {author} {\bibfnamefont {F.~A.}\ \bibnamefont {Ivanyuk}}, \bibinfo
  {author} {\bibfnamefont {J.~A.}\ \bibnamefont {Maruhn}}, \bibinfo {author}
  {\bibfnamefont {K.}~\bibnamefont {Nishio}}, \ and\ \bibinfo {author}
  {\bibfnamefont {S.}~\bibnamefont {Chiba}},\ }\href {\doibase
  10.1103/PhysRevC.96.064616} {\bibfield  {journal} {\bibinfo  {journal} {Phys.
  Rev. C}\ }\textbf {\bibinfo {volume} {96}},\ \bibinfo {pages} {064616}
  (\bibinfo {year} {2017})}\BibitemShut {NoStop}%
\bibitem [{\citenamefont {Ivanyuk}\ \emph {et~al.}(2024)\citenamefont
  {Ivanyuk}, \citenamefont {Ishizuka},\ and\ \citenamefont
  {Chiba}}]{ivanyuk20235dimensional}%
  \BibitemOpen
  \bibfield  {author} {\bibinfo {author} {\bibfnamefont {F.~A.}\ \bibnamefont
  {Ivanyuk}}, \bibinfo {author} {\bibfnamefont {C.}~\bibnamefont {Ishizuka}}, \
  and\ \bibinfo {author} {\bibfnamefont {S.}~\bibnamefont {Chiba}},\ }\href
  {\doibase 10.1103/PhysRevC.109.034602} {\bibfield  {journal} {\bibinfo
  {journal} {Phys. Rev. C}\ }\textbf {\bibinfo {volume} {109}},\ \bibinfo
  {pages} {034602} (\bibinfo {year} {2024})}\BibitemShut {NoStop}%
\bibitem [{\citenamefont {Huang}\ \emph {et~al.}(2022)\citenamefont {Huang},
  \citenamefont {Feng}, \citenamefont {Xiao}, \citenamefont {Lei},
  \citenamefont {Zhu},\ and\ \citenamefont {Su}}]{PhysRevC.106.054606}%
  \BibitemOpen
  \bibfield  {author} {\bibinfo {author} {\bibfnamefont {Y.}~\bibnamefont
  {Huang}}, \bibinfo {author} {\bibfnamefont {Y.}~\bibnamefont {Feng}},
  \bibinfo {author} {\bibfnamefont {E.}~\bibnamefont {Xiao}}, \bibinfo {author}
  {\bibfnamefont {X.}~\bibnamefont {Lei}}, \bibinfo {author} {\bibfnamefont
  {L.}~\bibnamefont {Zhu}}, \ and\ \bibinfo {author} {\bibfnamefont
  {J.}~\bibnamefont {Su}},\ }\href {\doibase 10.1103/PhysRevC.106.054606}
  {\bibfield  {journal} {\bibinfo  {journal} {Phys. Rev. C}\ }\textbf {\bibinfo
  {volume} {106}},\ \bibinfo {pages} {054606} (\bibinfo {year}
  {2022})}\BibitemShut {NoStop}%
\bibitem [{\citenamefont {Maruhn}\ and\ \citenamefont
  {Greiner}(1972)}]{MaruhnGreiner-96}%
  \BibitemOpen
  \bibfield  {author} {\bibinfo {author} {\bibfnamefont {J.}~\bibnamefont
  {Maruhn}}\ and\ \bibinfo {author} {\bibfnamefont {W.}~\bibnamefont
  {Greiner}},\ }\href {\doibase 10.1007/BF01391737} {\bibfield  {journal}
  {\bibinfo  {journal} {Z. Phys.}\ }\textbf {\bibinfo {volume} {251}},\
  \bibinfo {pages} {431} (\bibinfo {year} {1972})}\BibitemShut {NoStop}%
\bibitem [{\citenamefont {Aritomo}\ \emph {et~al.}(2014)\citenamefont
  {Aritomo}, \citenamefont {Chiba},\ and\ \citenamefont
  {Ivanyuk}}]{AritomoChiba-40}%
  \BibitemOpen
  \bibfield  {author} {\bibinfo {author} {\bibfnamefont {Y.}~\bibnamefont
  {Aritomo}}, \bibinfo {author} {\bibfnamefont {S.}~\bibnamefont {Chiba}}, \
  and\ \bibinfo {author} {\bibfnamefont {F.}~\bibnamefont {Ivanyuk}},\ }\href
  {\doibase 10.1103/PhysRevC.90.054609} {\bibfield  {journal} {\bibinfo
  {journal} {Phys. Rev. C}\ }\textbf {\bibinfo {volume} {90}},\ \bibinfo
  {pages} {054609} (\bibinfo {year} {2014})}\BibitemShut {NoStop}%
\bibitem [{\citenamefont {Ignatyuk}\ \emph {et~al.}(1979)\citenamefont
  {Ignatyuk}, \citenamefont {Istekov},\ and\ \citenamefont
  {Smirenkin}}]{ignatyuk1979}%
  \BibitemOpen
  \bibfield  {author} {\bibinfo {author} {\bibfnamefont {A.~V.}\ \bibnamefont
  {Ignatyuk}}, \bibinfo {author} {\bibfnamefont {K.~K.}\ \bibnamefont
  {Istekov}}, \ and\ \bibinfo {author} {\bibfnamefont {G.~N.}\ \bibnamefont
  {Smirenkin}},\ }\href@noop {} {\bibfield  {journal} {\bibinfo  {journal}
  {Sov. J. Nucl. Phys.}\ }\textbf {\bibinfo {volume} {29}},\ \bibinfo {pages}
  {450} (\bibinfo {year} {1979})}\BibitemShut {NoStop}%
\bibitem [{\citenamefont {Karpov}\ \emph {et~al.}(2017)\citenamefont {Karpov},
  \citenamefont {Denikin}, \citenamefont {Naumenko}, \citenamefont {Alekseev},
  \citenamefont {Rachkov}, \citenamefont {Samarin}, \citenamefont {Saiko},\
  and\ \citenamefont {Zagrebaev}}]{KARPOV2017112}%
  \BibitemOpen
  \bibfield  {author} {\bibinfo {author} {\bibfnamefont {A.}~\bibnamefont
  {Karpov}}, \bibinfo {author} {\bibfnamefont {A.}~\bibnamefont {Denikin}},
  \bibinfo {author} {\bibfnamefont {M.}~\bibnamefont {Naumenko}}, \bibinfo
  {author} {\bibfnamefont {A.}~\bibnamefont {Alekseev}}, \bibinfo {author}
  {\bibfnamefont {V.}~\bibnamefont {Rachkov}}, \bibinfo {author} {\bibfnamefont
  {V.}~\bibnamefont {Samarin}}, \bibinfo {author} {\bibfnamefont
  {V.}~\bibnamefont {Saiko}}, \ and\ \bibinfo {author} {\bibfnamefont
  {V.}~\bibnamefont {Zagrebaev}},\ }\href {\doibase
  https://doi.org/10.1016/j.nima.2017.01.069} {\bibfield  {journal} {\bibinfo
  {journal} {Nucl. Instrum. Methods. Phys. Res. A}\ }\textbf {\bibinfo {volume}
  {859}},\ \bibinfo {pages} {112} (\bibinfo {year} {2017})}\BibitemShut
  {NoStop}%
\bibitem [{\citenamefont {Krappe}\ \emph {et~al.}(1979)\citenamefont {Krappe},
  \citenamefont {Nix},\ and\ \citenamefont {Sierk}}]{PhysRevC.20.992}%
  \BibitemOpen
  \bibfield  {author} {\bibinfo {author} {\bibfnamefont {H.~J.}\ \bibnamefont
  {Krappe}}, \bibinfo {author} {\bibfnamefont {J.~R.}\ \bibnamefont {Nix}}, \
  and\ \bibinfo {author} {\bibfnamefont {A.~J.}\ \bibnamefont {Sierk}},\ }\href
  {\doibase 10.1103/PhysRevC.20.992} {\bibfield  {journal} {\bibinfo  {journal}
  {Phys. Rev. C}\ }\textbf {\bibinfo {volume} {20}},\ \bibinfo {pages} {992}
  (\bibinfo {year} {1979})}\BibitemShut {NoStop}%
\bibitem [{\citenamefont {Davies}\ and\ \citenamefont
  {Nix}(1976)}]{DaviesNix-31}%
  \BibitemOpen
  \bibfield  {author} {\bibinfo {author} {\bibfnamefont {K.~T.~R.}\
  \bibnamefont {Davies}}\ and\ \bibinfo {author} {\bibfnamefont {J.~R.}\
  \bibnamefont {Nix}},\ }\href {\doibase 10.1103/PhysRevC.14.1977} {\bibfield
  {journal} {\bibinfo  {journal} {Phys. Rev. C}\ }\textbf {\bibinfo {volume}
  {14}},\ \bibinfo {pages} {1977} (\bibinfo {year} {1976})}\BibitemShut
  {NoStop}%
\bibitem [{\citenamefont {Mavlitov}\ \emph {et~al.}(1992)\citenamefont
  {Mavlitov}, \citenamefont {Fr{\"o}brich},\ and\ \citenamefont
  {Gonchar}}]{mavlitov1992combining}%
  \BibitemOpen
  \bibfield  {author} {\bibinfo {author} {\bibfnamefont {N.~D.}\ \bibnamefont
  {Mavlitov}}, \bibinfo {author} {\bibfnamefont {P.}~\bibnamefont
  {Fr{\"o}brich}}, \ and\ \bibinfo {author} {\bibfnamefont {I.~I.}\
  \bibnamefont {Gonchar}},\ }\href@noop {} {\bibfield  {journal} {\bibinfo
  {journal} {Z. Phys. A}\ }\textbf {\bibinfo {volume} {342}},\ \bibinfo {pages}
  {195} (\bibinfo {year} {1992})}\BibitemShut {NoStop}%
\bibitem [{\citenamefont {Davies}\ \emph {et~al.}(1976)\citenamefont {Davies},
  \citenamefont {Sierk},\ and\ \citenamefont {Nix}}]{PhysRevC.13.2385}%
  \BibitemOpen
  \bibfield  {author} {\bibinfo {author} {\bibfnamefont {K.~T.~R.}\
  \bibnamefont {Davies}}, \bibinfo {author} {\bibfnamefont {A.~J.}\
  \bibnamefont {Sierk}}, \ and\ \bibinfo {author} {\bibfnamefont {J.~R.}\
  \bibnamefont {Nix}},\ }\href {\doibase 10.1103/PhysRevC.13.2385} {\bibfield
  {journal} {\bibinfo  {journal} {Phys. Rev. C}\ }\textbf {\bibinfo {volume}
  {13}},\ \bibinfo {pages} {2385} (\bibinfo {year} {1976})}\BibitemShut
  {NoStop}%
\bibitem [{\citenamefont {Krappe}\ and\ \citenamefont
  {Pomorski}(2012)}]{Krappe2012TheoryON}%
  \BibitemOpen
  \bibfield  {author} {\bibinfo {author} {\bibfnamefont {H.~J.}\ \bibnamefont
  {Krappe}}\ and\ \bibinfo {author} {\bibfnamefont {K.}~\bibnamefont
  {Pomorski}},\ }\href {\doibase https://doi.org/10.1007/978-3-642-23515-3}
  {\emph {\bibinfo {title} {Theory of Nuclear Fission: A Textbook}}},\ Vol.\
  \bibinfo {volume} {838}\ (\bibinfo  {publisher} {Springer
  Berlin/Heidelberg},\ \bibinfo {year} {2012})\BibitemShut {NoStop}%
\bibitem [{\citenamefont {Blocki}\ \emph {et~al.}(1978)\citenamefont {Blocki},
  \citenamefont {Boneh}, \citenamefont {Nix}, \citenamefont {Randrup},
  \citenamefont {Robel}, \citenamefont {Sierk},\ and\ \citenamefont
  {Swiatecki}}]{BlockiBoneh-22}%
  \BibitemOpen
  \bibfield  {author} {\bibinfo {author} {\bibfnamefont {J.}~\bibnamefont
  {Blocki}}, \bibinfo {author} {\bibfnamefont {Y.}~\bibnamefont {Boneh}},
  \bibinfo {author} {\bibfnamefont {J.}~\bibnamefont {Nix}}, \bibinfo {author}
  {\bibfnamefont {J.}~\bibnamefont {Randrup}}, \bibinfo {author} {\bibfnamefont
  {M.}~\bibnamefont {Robel}}, \bibinfo {author} {\bibfnamefont
  {A.}~\bibnamefont {Sierk}}, \ and\ \bibinfo {author} {\bibfnamefont
  {W.}~\bibnamefont {Swiatecki}},\ }\href {\doibase
  https://doi.org/10.1016/0003-4916(78)90208-7} {\bibfield  {journal} {\bibinfo
   {journal} {Ann. Phys.}\ }\textbf {\bibinfo {volume} {113}},\ \bibinfo
  {pages} {330} (\bibinfo {year} {1978})}\BibitemShut {NoStop}%
\bibitem [{\citenamefont {Sierk}\ and\ \citenamefont
  {Nix}(1980)}]{SierkNix-51}%
  \BibitemOpen
  \bibfield  {author} {\bibinfo {author} {\bibfnamefont {A.~J.}\ \bibnamefont
  {Sierk}}\ and\ \bibinfo {author} {\bibfnamefont {J.~R.}\ \bibnamefont
  {Nix}},\ }\href {\doibase 10.1103/PhysRevC.21.982} {\bibfield  {journal}
  {\bibinfo  {journal} {Phys. Rev. C}\ }\textbf {\bibinfo {volume} {21}},\
  \bibinfo {pages} {982} (\bibinfo {year} {1980})}\BibitemShut {NoStop}%
\bibitem [{\citenamefont {Adeev}\ \emph {et~al.}(2005)\citenamefont {Adeev},
  \citenamefont {Karpov}, \citenamefont {Nadtochy},\ and\ \citenamefont
  {Vanin}}]{AdeevVanin-53}%
  \BibitemOpen
  \bibfield  {author} {\bibinfo {author} {\bibfnamefont {G.~D.}\ \bibnamefont
  {Adeev}}, \bibinfo {author} {\bibfnamefont {A.~V.}\ \bibnamefont {Karpov}},
  \bibinfo {author} {\bibfnamefont {P.~N.}\ \bibnamefont {Nadtochy}}, \ and\
  \bibinfo {author} {\bibfnamefont {D.~V.}\ \bibnamefont {Vanin}},\ }\href@noop
  {} {\bibfield  {journal} {\bibinfo  {journal} {Phys. Part. Nucl.}\ }\textbf
  {\bibinfo {volume} {36}},\ \bibinfo {pages} {378} (\bibinfo {year}
  {2005})}\BibitemShut {NoStop}%
\bibitem [{\citenamefont {Swiatecki}(1984)}]{Swiatecki-52}%
  \BibitemOpen
  \bibfield  {author} {\bibinfo {author} {\bibfnamefont {W.~J.}\ \bibnamefont
  {Swiatecki}},\ }\href {\doibase https://doi.org/10.1016/0375-9474(84)90252-5}
  {\bibfield  {journal} {\bibinfo  {journal} {Nucl. Phys. A}\ }\textbf
  {\bibinfo {volume} {428}},\ \bibinfo {pages} {199} (\bibinfo {year}
  {1984})}\BibitemShut {NoStop}%
\bibitem [{\citenamefont {Rayford~Nix}\ and\ \citenamefont
  {Sierk}(1984)}]{RayfordNixSierk-35}%
  \BibitemOpen
  \bibfield  {author} {\bibinfo {author} {\bibfnamefont {J.}~\bibnamefont
  {Rayford~Nix}}\ and\ \bibinfo {author} {\bibfnamefont {A.~J.}\ \bibnamefont
  {Sierk}},\ }\href {\doibase https://doi.org/10.1016/0375-9474(84)90249-5}
  {\bibfield  {journal} {\bibinfo  {journal} {Nucl. Phys. A}\ }\textbf
  {\bibinfo {volume} {428}},\ \bibinfo {pages} {161} (\bibinfo {year}
  {1984})}\BibitemShut {NoStop}%
\bibitem [{\citenamefont {BŽocki}(1984)}]{Bzocki-37}%
  \BibitemOpen
  \bibfield  {author} {\bibinfo {author} {\bibfnamefont {J.}~\bibnamefont
  {BŽocki}},\ }\href {\doibase 10.1051/jphyscol:1984658} {\bibfield  {journal}
  {\bibinfo  {journal} {J. Phys. Colloq.}\ }\textbf {\bibinfo {volume} {45}},\
  \bibinfo {pages} {489} (\bibinfo {year} {1984})}\BibitemShut {NoStop}%
\bibitem [{\citenamefont {Feldmeier}(1987)}]{Feldmeier-36}%
  \BibitemOpen
  \bibfield  {author} {\bibinfo {author} {\bibfnamefont {H.}~\bibnamefont
  {Feldmeier}},\ }\href {\doibase 10.1088/0034-4885/50/8/001} {\bibfield
  {journal} {\bibinfo  {journal} {Rep. Prog. Phys.}\ }\textbf {\bibinfo
  {volume} {50}},\ \bibinfo {pages} {915} (\bibinfo {year} {1987})}\BibitemShut
  {NoStop}%
\bibitem [{\citenamefont {Viola}\ \emph {et~al.}(1985)\citenamefont {Viola},
  \citenamefont {Kwiatkowski},\ and\ \citenamefont
  {Walker}}]{PhysRevC.31.1550}%
  \BibitemOpen
  \bibfield  {author} {\bibinfo {author} {\bibfnamefont {V.~E.}\ \bibnamefont
  {Viola}}, \bibinfo {author} {\bibfnamefont {K.}~\bibnamefont {Kwiatkowski}},
  \ and\ \bibinfo {author} {\bibfnamefont {M.}~\bibnamefont {Walker}},\ }\href
  {\doibase 10.1103/PhysRevC.31.1550} {\bibfield  {journal} {\bibinfo
  {journal} {Phys. Rev. C}\ }\textbf {\bibinfo {volume} {31}},\ \bibinfo
  {pages} {1550} (\bibinfo {year} {1985})}\BibitemShut {NoStop}%
\bibitem [{\citenamefont {T\={o}ke}\ \emph {et~al.}(1985)\citenamefont
  {T\={o}ke}, \citenamefont {Bock}, \citenamefont {Dai}, \citenamefont {Gobbi},
  \citenamefont {Gralla}, \citenamefont {Hildenbrand}, \citenamefont
  {Kuzminski}, \citenamefont {Müller}, \citenamefont {Olmi}, \citenamefont
  {Stelzer} \emph {et~al.}}]{TOKE1985327}%
  \BibitemOpen
  \bibfield  {author} {\bibinfo {author} {\bibfnamefont {J.}~\bibnamefont
  {T\={o}ke}}, \bibinfo {author} {\bibfnamefont {R.}~\bibnamefont {Bock}},
  \bibinfo {author} {\bibfnamefont {G.}~\bibnamefont {Dai}}, \bibinfo {author}
  {\bibfnamefont {A.}~\bibnamefont {Gobbi}}, \bibinfo {author} {\bibfnamefont
  {S.}~\bibnamefont {Gralla}}, \bibinfo {author} {\bibfnamefont
  {K.}~\bibnamefont {Hildenbrand}}, \bibinfo {author} {\bibfnamefont
  {J.}~\bibnamefont {Kuzminski}}, \bibinfo {author} {\bibfnamefont
  {W.}~\bibnamefont {Müller}}, \bibinfo {author} {\bibfnamefont
  {A.}~\bibnamefont {Olmi}}, \bibinfo {author} {\bibfnamefont {H.}~\bibnamefont
  {Stelzer}},  \emph {et~al.},\ }\href {\doibase
  https://doi.org/10.1016/0375-9474(85)90344-6} {\bibfield  {journal} {\bibinfo
   {journal} {Nucl. Phys. A}\ }\textbf {\bibinfo {volume} {440}},\ \bibinfo
  {pages} {327} (\bibinfo {year} {1985})}\BibitemShut {NoStop}%
\bibitem [{\citenamefont {Nishio}\ \emph {et~al.}(2015)\citenamefont {Nishio},
  \citenamefont {Andreyev}, \citenamefont {Chapman}, \citenamefont {Derkx},
  \citenamefont {Düllmann}, \citenamefont {Ghys}, \citenamefont {Heßberger},
  \citenamefont {Hirose}, \citenamefont {Ikezoe}, \citenamefont {Khuyagbaatar}
  \emph {et~al.}}]{NISHIO201589}%
  \BibitemOpen
  \bibfield  {author} {\bibinfo {author} {\bibfnamefont {K.}~\bibnamefont
  {Nishio}}, \bibinfo {author} {\bibfnamefont {A.}~\bibnamefont {Andreyev}},
  \bibinfo {author} {\bibfnamefont {R.}~\bibnamefont {Chapman}}, \bibinfo
  {author} {\bibfnamefont {X.}~\bibnamefont {Derkx}}, \bibinfo {author}
  {\bibfnamefont {C.}~\bibnamefont {Düllmann}}, \bibinfo {author}
  {\bibfnamefont {L.}~\bibnamefont {Ghys}}, \bibinfo {author} {\bibfnamefont
  {F.}~\bibnamefont {Heßberger}}, \bibinfo {author} {\bibfnamefont
  {K.}~\bibnamefont {Hirose}}, \bibinfo {author} {\bibfnamefont
  {H.}~\bibnamefont {Ikezoe}}, \bibinfo {author} {\bibfnamefont
  {J.}~\bibnamefont {Khuyagbaatar}},  \emph {et~al.},\ }\href {\doibase
  https://doi.org/10.1016/j.physletb.2015.06.068} {\bibfield  {journal}
  {\bibinfo  {journal} {Phys. Lett. B}\ }\textbf {\bibinfo {volume} {748}},\
  \bibinfo {pages} {89} (\bibinfo {year} {2015})}\BibitemShut {NoStop}%
\bibitem [{\citenamefont {Prasad}\ \emph {et~al.}(2020)\citenamefont {Prasad},
  \citenamefont {Hinde}, \citenamefont {Dasgupta}, \citenamefont {Jeung},
  \citenamefont {Berriman}, \citenamefont {Swinton-Bland}, \citenamefont
  {Simenel}, \citenamefont {Simpson}, \citenamefont {Bernard}, \citenamefont
  {Williams} \emph {et~al.}}]{PRASAD2020135941}%
  \BibitemOpen
  \bibfield  {author} {\bibinfo {author} {\bibfnamefont {E.}~\bibnamefont
  {Prasad}}, \bibinfo {author} {\bibfnamefont {D.}~\bibnamefont {Hinde}},
  \bibinfo {author} {\bibfnamefont {M.}~\bibnamefont {Dasgupta}}, \bibinfo
  {author} {\bibfnamefont {D.}~\bibnamefont {Jeung}}, \bibinfo {author}
  {\bibfnamefont {A.}~\bibnamefont {Berriman}}, \bibinfo {author}
  {\bibfnamefont {B.}~\bibnamefont {Swinton-Bland}}, \bibinfo {author}
  {\bibfnamefont {C.}~\bibnamefont {Simenel}}, \bibinfo {author} {\bibfnamefont
  {E.}~\bibnamefont {Simpson}}, \bibinfo {author} {\bibfnamefont
  {R.}~\bibnamefont {Bernard}}, \bibinfo {author} {\bibfnamefont
  {E.}~\bibnamefont {Williams}},  \emph {et~al.},\ }\href {\doibase
  https://doi.org/10.1016/j.physletb.2020.135941} {\bibfield  {journal}
  {\bibinfo  {journal} {Phys. Lett. B}\ }\textbf {\bibinfo {volume} {811}},\
  \bibinfo {pages} {135941} (\bibinfo {year} {2020})}\BibitemShut {NoStop}%
\bibitem [{\citenamefont {Tsekhanovich}\ \emph {et~al.}(2019)\citenamefont
  {Tsekhanovich}, \citenamefont {Andreyev}, \citenamefont {Nishio},
  \citenamefont {Denis-Petit}, \citenamefont {Hirose}, \citenamefont {Makii},
  \citenamefont {Matheson}, \citenamefont {Morimoto}, \citenamefont {Morita},
  \citenamefont {Nazarewicz} \emph {et~al.}}]{TSEKHANOVICH2019583}%
  \BibitemOpen
  \bibfield  {author} {\bibinfo {author} {\bibfnamefont {I.}~\bibnamefont
  {Tsekhanovich}}, \bibinfo {author} {\bibfnamefont {A.}~\bibnamefont
  {Andreyev}}, \bibinfo {author} {\bibfnamefont {K.}~\bibnamefont {Nishio}},
  \bibinfo {author} {\bibfnamefont {D.}~\bibnamefont {Denis-Petit}}, \bibinfo
  {author} {\bibfnamefont {K.}~\bibnamefont {Hirose}}, \bibinfo {author}
  {\bibfnamefont {H.}~\bibnamefont {Makii}}, \bibinfo {author} {\bibfnamefont
  {Z.}~\bibnamefont {Matheson}}, \bibinfo {author} {\bibfnamefont
  {K.}~\bibnamefont {Morimoto}}, \bibinfo {author} {\bibfnamefont
  {K.}~\bibnamefont {Morita}}, \bibinfo {author} {\bibfnamefont
  {W.}~\bibnamefont {Nazarewicz}},  \emph {et~al.},\ }\href {\doibase
  https://doi.org/10.1016/j.physletb.2019.02.006} {\bibfield  {journal}
  {\bibinfo  {journal} {Phys. Lett. B}\ }\textbf {\bibinfo {volume} {790}},\
  \bibinfo {pages} {583} (\bibinfo {year} {2019})}\BibitemShut {NoStop}%
\bibitem [{\citenamefont {Bogachev}\ \emph {et~al.}(2021)\citenamefont
  {Bogachev}, \citenamefont {Kozulin}, \citenamefont {Knyazheva}, \citenamefont
  {Itkis}, \citenamefont {Itkis}, \citenamefont {Novikov}, \citenamefont
  {Kumar}, \citenamefont {Banerjee}, \citenamefont {Diatlov}, \citenamefont
  {Cheralu} \emph {et~al.}}]{PhysRevC.104.024623}%
  \BibitemOpen
  \bibfield  {author} {\bibinfo {author} {\bibfnamefont {A.~A.}\ \bibnamefont
  {Bogachev}}, \bibinfo {author} {\bibfnamefont {E.~M.}\ \bibnamefont
  {Kozulin}}, \bibinfo {author} {\bibfnamefont {G.~N.}\ \bibnamefont
  {Knyazheva}}, \bibinfo {author} {\bibfnamefont {I.~M.}\ \bibnamefont
  {Itkis}}, \bibinfo {author} {\bibfnamefont {M.~G.}\ \bibnamefont {Itkis}},
  \bibinfo {author} {\bibfnamefont {K.~V.}\ \bibnamefont {Novikov}}, \bibinfo
  {author} {\bibfnamefont {D.}~\bibnamefont {Kumar}}, \bibinfo {author}
  {\bibfnamefont {T.}~\bibnamefont {Banerjee}}, \bibinfo {author}
  {\bibfnamefont {I.~N.}\ \bibnamefont {Diatlov}}, \bibinfo {author}
  {\bibfnamefont {M.}~\bibnamefont {Cheralu}},  \emph {et~al.},\ }\href
  {\doibase 10.1103/PhysRevC.104.024623} {\bibfield  {journal} {\bibinfo
  {journal} {Phys. Rev. C}\ }\textbf {\bibinfo {volume} {104}},\ \bibinfo
  {pages} {024623} (\bibinfo {year} {2021})}\BibitemShut {NoStop}%
\bibitem [{\citenamefont {Kozulin}\ \emph {et~al.}(2022)\citenamefont
  {Kozulin}, \citenamefont {Knyazheva}, \citenamefont {Itkis}, \citenamefont
  {Itkis}, \citenamefont {Mukhamejanov}, \citenamefont {Bogachev},
  \citenamefont {Novikov}, \citenamefont {Kirakosyan}, \citenamefont {Kumar},
  \citenamefont {Banerjee} \emph {et~al.}}]{PhysRevC.105.014607}%
  \BibitemOpen
  \bibfield  {author} {\bibinfo {author} {\bibfnamefont {E.~M.}\ \bibnamefont
  {Kozulin}}, \bibinfo {author} {\bibfnamefont {G.~N.}\ \bibnamefont
  {Knyazheva}}, \bibinfo {author} {\bibfnamefont {I.~M.}\ \bibnamefont
  {Itkis}}, \bibinfo {author} {\bibfnamefont {M.~G.}\ \bibnamefont {Itkis}},
  \bibinfo {author} {\bibfnamefont {Y.~S.}\ \bibnamefont {Mukhamejanov}},
  \bibinfo {author} {\bibfnamefont {A.~A.}\ \bibnamefont {Bogachev}}, \bibinfo
  {author} {\bibfnamefont {K.~V.}\ \bibnamefont {Novikov}}, \bibinfo {author}
  {\bibfnamefont {V.~V.}\ \bibnamefont {Kirakosyan}}, \bibinfo {author}
  {\bibfnamefont {D.}~\bibnamefont {Kumar}}, \bibinfo {author} {\bibfnamefont
  {T.}~\bibnamefont {Banerjee}},  \emph {et~al.},\ }\href {\doibase
  10.1103/PhysRevC.105.014607} {\bibfield  {journal} {\bibinfo  {journal}
  {Phys. Rev. C}\ }\textbf {\bibinfo {volume} {105}},\ \bibinfo {pages}
  {014607} (\bibinfo {year} {2022})}\BibitemShut {NoStop}%
\bibitem [{\citenamefont {Kumar}\ \emph {et~al.}(2023)\citenamefont {Kumar},
  \citenamefont {Maiti}, \citenamefont {Pal}, \citenamefont {Santra},
  \citenamefont {Kaur}, \citenamefont {Sagwal}, \citenamefont {Singh},
  \citenamefont {Rout}, \citenamefont {Baishya}, \citenamefont {Gandhi},\ and\
  \citenamefont {Santhosh}}]{PhysRevC.107.034614}%
  \BibitemOpen
  \bibfield  {author} {\bibinfo {author} {\bibfnamefont {R.}~\bibnamefont
  {Kumar}}, \bibinfo {author} {\bibfnamefont {M.}~\bibnamefont {Maiti}},
  \bibinfo {author} {\bibfnamefont {A.}~\bibnamefont {Pal}}, \bibinfo {author}
  {\bibfnamefont {S.}~\bibnamefont {Santra}}, \bibinfo {author} {\bibfnamefont
  {P.}~\bibnamefont {Kaur}}, \bibinfo {author} {\bibfnamefont {M.}~\bibnamefont
  {Sagwal}}, \bibinfo {author} {\bibfnamefont {A.}~\bibnamefont {Singh}},
  \bibinfo {author} {\bibfnamefont {P.~C.}\ \bibnamefont {Rout}}, \bibinfo
  {author} {\bibfnamefont {A.}~\bibnamefont {Baishya}}, \bibinfo {author}
  {\bibfnamefont {R.}~\bibnamefont {Gandhi}}, \ and\ \bibinfo {author}
  {\bibfnamefont {T.}~\bibnamefont {Santhosh}},\ }\href {\doibase
  10.1103/PhysRevC.107.034614} {\bibfield  {journal} {\bibinfo  {journal}
  {Phys. Rev. C}\ }\textbf {\bibinfo {volume} {107}},\ \bibinfo {pages}
  {034614} (\bibinfo {year} {2023})}\BibitemShut {NoStop}%
\bibitem [{\citenamefont {Miernik}\ \emph {et~al.}(2023)\citenamefont
  {Miernik}, \citenamefont {Korgul}, \citenamefont {Poklepa}, \citenamefont
  {Wilson}, \citenamefont {Charles}, \citenamefont {Czajkowski}, \citenamefont
  {Czy\ifmmode~\dot{z}\else \.{z}\fi{}}, \citenamefont {Fija\l{}kowska},
  \citenamefont {Fraile}, \citenamefont {Garczy\ifmmode~\acute{n}\else
  \'{n}\fi{}ski} \emph {et~al.}}]{PhysRevC.108.054608}%
  \BibitemOpen
  \bibfield  {author} {\bibinfo {author} {\bibfnamefont {K.}~\bibnamefont
  {Miernik}}, \bibinfo {author} {\bibfnamefont {A.}~\bibnamefont {Korgul}},
  \bibinfo {author} {\bibfnamefont {W.}~\bibnamefont {Poklepa}}, \bibinfo
  {author} {\bibfnamefont {J.~N.}\ \bibnamefont {Wilson}}, \bibinfo {author}
  {\bibfnamefont {G.}~\bibnamefont {Charles}}, \bibinfo {author} {\bibfnamefont
  {S.}~\bibnamefont {Czajkowski}}, \bibinfo {author} {\bibfnamefont
  {P.}~\bibnamefont {Czy\ifmmode~\dot{z}\else \.{z}\fi{}}}, \bibinfo {author}
  {\bibfnamefont {A.}~\bibnamefont {Fija\l{}kowska}}, \bibinfo {author}
  {\bibfnamefont {L.~M.}\ \bibnamefont {Fraile}}, \bibinfo {author}
  {\bibfnamefont {P.}~\bibnamefont {Garczy\ifmmode~\acute{n}\else
  \'{n}\fi{}ski}},  \emph {et~al.},\ }\href {\doibase
  10.1103/PhysRevC.108.054608} {\bibfield  {journal} {\bibinfo  {journal}
  {Phys. Rev. C}\ }\textbf {\bibinfo {volume} {108}},\ \bibinfo {pages}
  {054608} (\bibinfo {year} {2023})}\BibitemShut {NoStop}%
\end{thebibliography}%

\end{document}